\documentclass[journal]{IEEEtran}
\usepackage{cite}

\ifCLASSINFOpdf
\else
\fi
\ifCLASSOPTIONcompsoc
  \usepackage[caption=false,font=normalsize,labelfont=sf,textfont=sf]{subfig}
\else
  \usepackage[caption=false,font=footnotesize]{subfig}
\fi
\usepackage{float}
\usepackage{amsfonts,amsopn,dsfont} 
\usepackage{color}
\usepackage{amsmath}
\usepackage[table]{xcolor}
\usepackage{graphicx,epstopdf}
\usepackage{hyperref}
\usepackage{cleveref}
\usepackage{algorithm}
\usepackage{algorithmic}
\usepackage{multirow}

\DeclareMathOperator*{\argmax}{arg\,max}
\DeclareMathOperator*{\argmin}{arg\,min}
\DeclareMathOperator*{\card}{\#}
\newcommand{\mat}[1]{\boldsymbol{#1}}
\newcommand{\myvec}[1]{\boldsymbol{#1}}
\newcommand{\indicator}[2]{\mathds{1}_{#1}(#2)}

\Crefformat{figure}{#2Fig.~#1#3}
\Crefmultiformat{figure}{Figs.~#2#1#3}{ and~#2#1#3}{, #2#1#3}{ and~#2#1#3}
\begin{document}
%
\title{Bayesian 3D Reconstruction of Subsampled Multispectral Single-photon Lidar Signals}
%
%
%

\author{Juli\'an~Tachella,~\IEEEmembership{Student Member,~IEEE,}
	Yoann~Altmann,~\IEEEmembership{Member,~IEEE,}
	Miguel M\'arquez,~\IEEEmembership{Student Member,~IEEE,} Henry Arguello-Fuentes,~\IEEEmembership{Senior Member,~IEEE,}
	Jean-Yves~Tourneret,~\IEEEmembership{Fellow,~IEEE}
	and~Stephen~McLaughlin,~\IEEEmembership{Fellow,~IEEE}
	\thanks{\textcopyright 2019 IEEE.  Personal use of this material is permitted.  Permission from IEEE must be obtained for all other uses, in any current or future media, including reprinting/republishing this material for advertising or promotional purposes, creating new collective works, for resale or redistribution to servers or lists, or reuse of any copyrighted component of this work in other works.  J. Tachella, Y. Altmann and S. McLaughlin are with the School of Engineering, Heriot-Watt University. M. M\'arquez is with the Department of Physics, Universidad Industrial de Santander. H. Arguello-Fuentes is with the Department of Systems Engineering, Universidad Industrial de Santander. J.-Y. Tourneret is with the Signal and Image Department, University of Toulouse. This work was supported by the Royal Academy of Engineering under
	the Research Fellowship scheme RF201617/16/31. This work was supported by EPSRC grant number EP/R033013/1. This work was partly conducted within the ECOS project 'Colored apertude design for compressive spectral imaging' supported by CNRS and Colciencias, and within the STIC-AmSud Project HyperMed. MATLAB codes can be found at \url{https://gitlab.com/Tachella/musapop}.}}

\markboth{}%
{Tachella \MakeLowercase{\textit{et al.}}: {Bayesian 3D Reconstruction of Subsampled Multispectral Single-photon Lidar Signals}}
%



\maketitle

\begin{abstract}
Light detection and ranging (Lidar) single-photon devices capture range and intensity information from a 3D scene. This modality enables long range 3D reconstruction with high range precision and low laser power.  A multispectral single-photon Lidar system provides additional spectral diversity, allowing the discrimination of different materials.  However, the main drawback of such systems can be the long acquisition time needed to collect enough photons in each spectral band. In this work, we tackle this problem in two ways: first, we propose a Bayesian 3D reconstruction algorithm that is able to find multiple surfaces per pixel, using few photons, i.e., shorter acquisitions. In contrast to previous algorithms, the novel method processes jointly all the spectral bands, obtaining better reconstructions using less photon detections. The proposed model promotes spatial correlation between neighbouring points within a given surface using spatial point processes. Secondly, we account for different spatial and spectral subsampling schemes, which reduce the total number of measurements, without significant degradation of the reconstruction performance. In this way, the total acquisition time, memory requirements and computational time can be significantly reduced. The experiments performed using both synthetic and real single-photon Lidar data demonstrate the advantages of tailored sampling schemes over random alternatives. Furthermore, the proposed algorithm yields better estimates than other existing methods for multi-surface reconstruction using multispectral Lidar data.

\end{abstract}


%
\IEEEpeerreviewmaketitle

\section{Introduction}
\begin{figure}[!h]
	\centering
	\includegraphics[width=.4\linewidth]{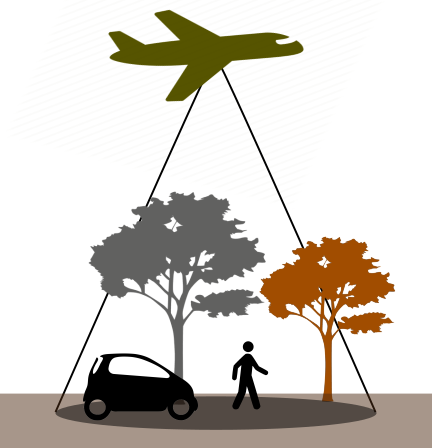}
	\caption{\textcolor{black}{An airborne MSL system can capture multiple objects per pixel and discriminate their materials. The multi-depth capability enables the recovery of information from photons reflected off different branches of the trees, ground, pedestrians or even from the interior of a car (i.e., photons which propagated across the windshield) at an intra-pixel level. Moreover, the multispectral information allows us to discriminate different properties of the materials of each 3D object (e.g., the leaves and trunk of a tree).}} 
	\label{FIG:app_msl}
\end{figure}

\begin{figure*}[!h]
	\centering
	\includegraphics[width=.8\textwidth]{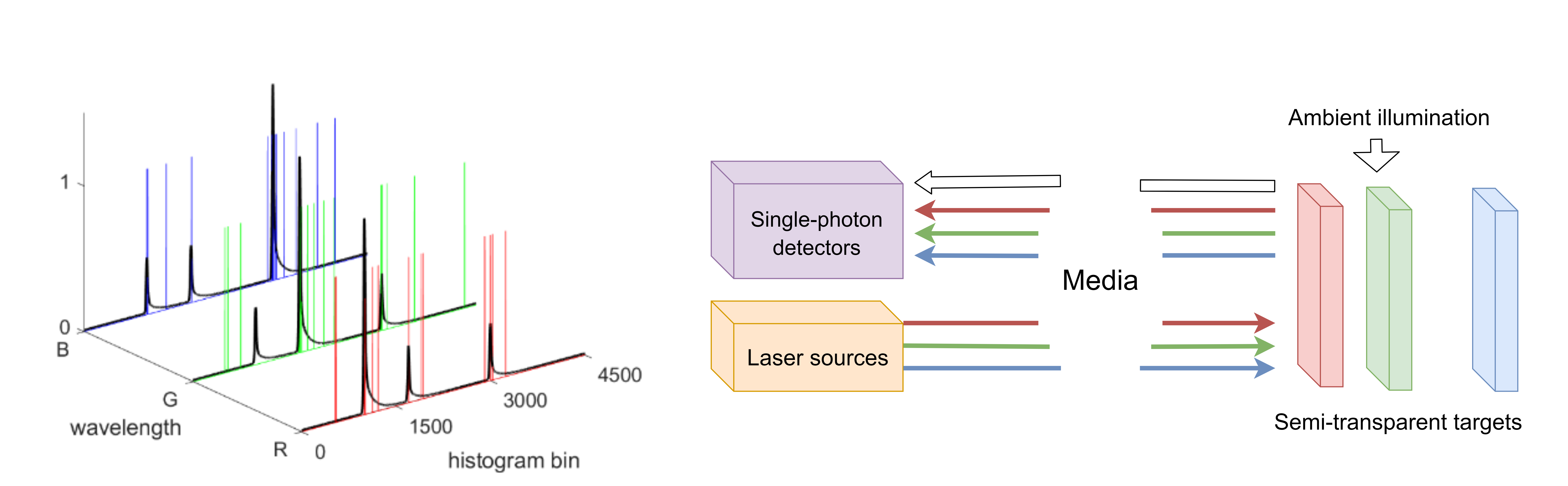}
	\caption{Example of an MSL system with three different wavelengths (red, green and blue). On the right, a schematic shows the working principles of a single-photon multispectral Lidar device. The red, green and blue arrows illustrate the laser pulses sent by the laser sources and reflected by the target onto the single-photon detectors. The white arrow depicts the background photons emitted by an ambient illumination source that reach the detectors at random times. The figure on the left shows the collected histograms for a given pixel: the discrete measurements are depicted in red, green and blue, while the underlying Poisson intensity (i.e., the parameters to estimate from the data) is shown in black.} 
	\label{FIG:example_msl}
\end{figure*}

Single-photon Lidar devices provide accurate range information by constructing, for each pixel, a histogram of time delays between emitted light pulses and photon detections. Using time correlated single-photon counting (TCSPC) technology, Lidar systems are able to provide accurate depth information (of the order of centimetres) over very long ranges, while allowing the use of laser sources of low power \cite{McCarthy:13}. The acquired range information (3D structure) has many important applications, such as self-driving cars \cite{hecht2018lidar}, the study of structures behind dense forests \cite{Canutoeaau0137} and environmental monitoring \cite{mallet2009full}. \textcolor{black}{Multispectral Lidar (MSL) systems gather measurements at many spectral bands, making it possible to distinguish distinct materials, as illustrated in \Cref{FIG:app_msl}. For example, spectral diversity was used in \cite{douglas2015echidna} to differentiate leaves from trunks and in \cite{wallace2014msl} to estimate plant area indices and abundance profiles. The MSL modality consists of constructing one histogram of time delays per wavelength, as shown in \Cref{FIG:example_msl}.} The spectral diversity can be obtained either using a supercontinuum laser source \cite{wallace2014msl,altmann2015msl} or multiple lasers \cite{WEI20121}. The detector generally consists of a spatial form of wavelength routing to demultiplex the channels \cite{wallace2014msl,altmann2015msl,WEI20121} or wavelength-to-time codification \cite{Ren:18}. Recovering spatial and spectral information from MSL data is a challenging task, specially in scenes with strong ambient illumination (i.e., multiple spurious detections) or when the acquisition time is very low (i.e., very few photon detections per histogram). Moreover, in a general setting, it is possible to find more than one object per pixel. This phenomenon occurs in scenes where the light goes through semi-transparent materials (e.g., glass), or when the laser footprint is such that multiple surfaces appear in the field of view. Thus, several signal processing algorithms have been proposed to address these challenges: while many algorithms are available for single-wavelength Lidar, either assuming a single surface per pixel \cite{kirmani2014first,Rapp2017AFP,Lindell:2018:3D} or multiple surfaces per pixel \cite{shin2016computational,halimi2016restoration,tachella2018manipop}, to the best of our knowledge, only the single-surface-per-pixel case was studied in the multispectral case \cite{altmann2015msl,altmann2017undersample,altmann2017robust}. 
\textcolor{black}{
This single-depth assumption greatly simplifies the reconstruction problem, as it significantly reduces memory requirements and overall complexity. Datasets containing dozens of wavelengths can be prohibitively large for practical 3D reconstruction algorithms, both in terms of memory and computing requirements. For example, a typical MSL hypercube with 32 wavelengths has more than $10^9$ data voxels. To alleviate this problem, some compressive acquisition strategies have been proposed. While TCSPC technology hinders compressive techniques along the depth axis\footnote{\textcolor{black}{A coarse time-of-flight gating is applied to the photon detections, hindering measurements of an arbitrary subset of histogram bins or linear combination of them. However, other alternatives such as gated cameras \cite{Ren18gated} can provide such measurements.}}, reducing the number of measurements can be achieved by  integrating multiple wavelengths in a single histogram \cite{Ren:18} or measuring fewer histograms (i.e., subsampling) \cite{altmann2017undersample}. The wavelength-to-time approach proposed by Ren et al. \cite{Ren:18} is not well-suited in the presence of multiple surfaces per pixel. Indeed, this method compresses $L$ histograms (associated with $L$ wavelengths) into a single waveform by shifting in time the photon detections according to each measured wavelength. While significantly reducing the data size, the resulting likelihood becomes highly multimodal and extremely difficult to handle in the presence of multiple surfaces. Different random subsampling schemes were studied in \cite{altmann2017undersample} without obtaining any significant differences in terms of reconstruction quality in the low-photon count regime.}

\textcolor{black}{In this work, we investigate a new pseudo-random subsampling scheme for low-photon count MSL data based on ideas from coded aperture design \cite{arguello2014colored,correa2016spatiotemporal}. By choosing the pixels measured for each wavelength in a more principled way, we achieve better results than the completely random schemes of Altmann et al. \cite{altmann2017undersample}. Furthermore, the proposed subsampling strategy can be easily implemented in many Lidar systems, reducing the total number of measurements, i.e., the time to acquire an MSL frame. Raster-scan systems using a laser supercontinuum source \cite{wallace2014msl,altmann2015msl} can be easily modified to measure only a subset of pixels per wavelength. More interestingly, single-wavelength array technology \cite{shin2016photon} can be combined with coded apertures \cite{arguello2014colored}, which acquire different wavelengths at each pixel.}

\textcolor{black}{Furthermore, we propose a new method to perform 3D reconstruction from subsampled MSL data, which is capable of finding multiple surfaces per pixel. The novel method draws ideas from a recently published algorithm named ManiPoP \cite{tachella2018manipop}, provided state-of-the-art reconstructions in the multiple surface, single-wavelength case. However, due to the significantly larger dimensionality of multispectral data, we propose to modify the Bayesian model and estimation strategy of ManiPoP. Adopting a Bayesian framework similar to \cite{tachella2018manipop}, we assign a spatial point process prior to promote spatial correlation and a Gaussian Markov random field prior to regularize the spectral reflectivity within surfaces/objects. Inference using the resulting posterior distribution is performed using a reversible jump Markov chain Monte Carlo algorithm (RJ-MCMC), coupled with a multiresolution approach that improves the convergence speed and reduces the total computing time of the algorithm. We introduce new RJ-MCMC proposals, which take into account the additional spectral dimension and improve the acceptance ratio, compared to the ones proposed in \cite{tachella2018manipop}.  Moreover, we propose an empirical Bayes approach to build the prior distribution associated with the background detections, which further improves the convergence of the RJ-MCMC sampler. Contrary to multi-depth methods that require storage of dense volumetric estimates (e.g., \cite{shin2016computational,halimi2016restoration} in the single-wavelength case and \cite{halimi2019mnrd} in the multi-temporal case), the memory requirements of the proposed method are minimal (just the 3D points and spectral signatures are stored in memory), enabling the acquisition and processing of very large datasets (dozens of wavelengths and hundreds of pixels).}

The main contributions of this work are
\begin{itemize}
	\item a new Bayesian 3D reconstruction algorithm for subsampled MSL data with multiple surfaces per pixel
	\item the analysis of an appropriate subsampling scheme based on coded aperture design, which provides better results than completely random alternatives.
\end{itemize}
The new algorithm is referred to as MuSaPoP, as it models MultiSpectrAl Lidar signals using POint Processes. The remainder of the paper is organized as follows. \Cref{SEC:MSL} recalls the classical observation model for single-photon MSL data. \Cref{SEC:MuMaPoP,SEC:RJ-MCMC} present the Bayesian 3D reconstruction algorithm and the associated RJ-MCMC sampler. \Cref{SEC:CS} details the principled subsampling strategy. Experiments performed with synthetic and real Lidar data are introduced and discussed in  \Cref{SEC:Exper}. Conclusions and future work are finally reported in \Cref{SEC:Conclusions}.

\section{Single-photon multispectral Lidar}\label{SEC:MSL}
A full multispectral Lidar data hypercube  $\mat{Z}\in\mathbb{Z}_{+}^{N_r\times N_c\times L \times T}$ consists of discrete photon count measurements, where $\mathbb{Z}_{+}=\{0,1,\dots\}$ is the set of positive integers, $N_r$ and $N_c$ are the numbers of vertical and horizontal pixels respectively, $L$ is the number of acquired wavelengths and $T$ is the histogram length (i.e., the number of time bins). The 3D reconstruction algorithm estimates a point cloud, referred to as  $\mat{\Phi}$, from the data hypercube $\mat{Z}$. The 3D point cloud is represented by an unordered set of points, that is
\begin{equation}\label{EQ:set_of_3D_points}
\mat{\Phi}=\{(\myvec{c}_n,\myvec{r}_n), n=1,\dots,N_\Phi\}
\end{equation}
where $N_\Phi$ is the total number of points, and $\myvec{c}_n = [x_n,y_n,t_n]^T \in [1,N_r]\times[1,N_c]\times[1,T]\subset\mathbb{Z}_{+}^{3}$ and $\myvec{r}_n = [r_{n,1},\dots,r_{n,L}]^T\in \mathbb{R}_{+}^L$ are the coordinate vector and the spectral response of the $n$th point, respectively. \textcolor{black}{T}he observed photon count at pixel $(i,j)$, bin $t$ and spectral band $\ell$ follows a Poisson distribution \cite{altmann2017undersample}, whose intensity is a mixture of the background level, denoted by $b_{i,j,\ell}$, and the responses of the surfaces present in that pixel, i.e.,
\begin{equation*}
\label{EQ:likelihood}
z_{i,j,\ell,t} | \mat{\Phi} ,b_{i,j,\ell}\sim \mathcal{P} \left( g_{i,j,\ell} \left(\sum_{\mathcal{N}_{i,j}}r_{n,\ell}h_{\ell}(t-t_{n})+b_{i,j,\ell} \right) \right)
\end{equation*}
where $\mathcal{P}(\cdot)$ denotes the Poisson distribution, $g_{i,j,\ell}\in \{0,1 \}$ is a known binary variable that indicates whether wavelength $\ell$ at pixel $(i,j)$ has been acquired/observed or not, $\mathcal{N}_{i,j}=\{n:(x_n,y_n)=(i,j)\}$ is the set of points corresponding to pixel $(i,j)$ and wavelength $\ell$, and $h_{\ell}(t)$ is the impulse response of the Lidar device at wavelength $\ell$, which can be measured using a spectralon panel during a calibration step. Note that to lighten notations, we assume that the impulse responses are only wavelength-dependent but the algorithm can be easily adapted when the responses vary among pixels. Assuming mutual independence between noise realizations in different bins, pixels and spectral bands \cite{McCarthy:13}, the likelihood of the proposed model can be written as
\begin{equation}
\label{EQ:full_likelihood}
p(\mat{Z} | \mat{\Phi},\mat{B}) = \prod_{i=1}^{N_c} \prod_{j=1}^{N_r} \prod_{\ell=1}^{L} \prod_{t=1}^{T} p(z_{i,j,\ell,t} | \mat{\Phi},b_{i,j,\ell}).
\end{equation}

For clarity in the notation, we will also denote the set of point coordinates as $
\mat{\Phi}_c=\{\myvec{c}_n, n=1,\dots,N_\Phi\}$ and the set of spectral responses as $\mat{\Phi}_r=\{\myvec{r}_n, n=1,\dots,N_\Phi\}$. The set of all background levels is denoted by $\mat{B} = [\myvec{b}_1,\dots,\myvec{b}_L] \in \mathbb{R}_{+}^{N_r\times N_c\times L}$, which is the concatenation of $L$ images $\myvec{b}_{\ell}$, one for each wavelength. The cube of binary measurements is designated by $\mat{G}\in\{0,1\}^{N_r\times N_c\times L}$, where $[\mat{G}]_{i,j,\ell} = g_{i,j,\ell}$. Note that the model used in the ManiPoP algorithm \cite{tachella2018manipop} can be obtained from \eqref{EQ:full_likelihood} by setting all the binary variables to 1, and considering only one band, i.e., $L=1$.

\section{Multiple-return multiple-wavelength 3D reconstruction}\label{SEC:MuMaPoP}
Recovering the position of the objects $\myvec{c}_n$, their spectral signatures $\myvec{r}_n$ and background levels $\mat{B}$ from the subsampled MSL data $\mat{Z}$ is an ill-posed inverse problem, as many solutions can explain the observed photon counts. Thus, prior regularization is necessary to promote reconstructions in a set of feasible 3D point clouds. In this work, we adopt a Bayesian framework, which allows us to include prior knowledge about the scene through tailored prior distributions assigned to the parameters of interest.

\subsection{Prior distributions}
The proposed model considers prior regularization for the point positions and reflectivity.  As explained in \Cref{SUBSEC:bkg_levels}, an empirical Bayes prior \cite{robert2007bayesian} is assigned to the background levels.
\subsubsection{Spatial configuration}
We adopt the spatial prior distribution of 3D points developed in the ManiPoP algorithm. This distribution is designed to promote spatial correlation between points within one surface and repulsion between points belonging to different surfaces. While only a brief summary is included below, we refer the reader to \cite{tachella2018manipop} for a detailed discussion about this prior model. The spatial point process prior for the position of the points is modelled by a density defined with respect to a Poisson point process reference measure \cite[Chapter~9]{brooks2011handbook}, i.e., 
\begin{equation*}
f(\mat{\Phi}_c) \propto f_1(\mat{\Phi}_c) f_2(\mat{\Phi}_c|\gamma_{a},\lambda_{a})
\end{equation*}
where $f_1(\mat{\Phi}_c)$ and $f_2(\mat{\Phi}_c|\gamma_{a},\lambda_{a})$ are the Strauss and area interaction \cite{baddeley1995area} processes respectively. The repulsive Strauss process is written as
\begin{equation*}
f_1(\mat{\Phi}_c) \propto \left\{
\begin{array}{ll}
0  & \text{if } \exists~ n \ne n' : x_n=x_{n'}, y_n=y_{n'} \\
&   \text{and } |t_{n}-t_{n'}| < d_{\min} \\
1 & \mbox{otherwise}
\end{array}
\right. 
\end{equation*}
where $d_{\min}$ is the minimum separation between two surfaces in the same pixel. Attraction between points within the same surface is promoted by the area interaction process, that is 
\begin{equation}
\label{EQ:area_interaction_density}
f_2(\mat{\Phi}_c|\gamma_{a},\lambda_{a}) \propto \lambda_{a}^{N_\Phi}\gamma_{a}^{-m\left( \bigcup_{n=1}^{N_\Phi}S(\myvec{c}_{n}) \right)}
\end{equation}
where $m(\cdot)$ denotes the standard Lebesgue measure, $S(\myvec{c}_{n})$ defines a convex set around the point $\myvec{c}_{n}$, and  $\gamma_{a}$ and $\lambda_{a}$ are two hyperparameters, accounting for the amount of attraction and total number of points, respectively. Both densities define Markovian interactions between points, only correlating points in a local neighbourhood.
Moreover, the combination of both processes implicitly defines a connected-surface structure, which is used to model 2D manifolds in a 3D space.
\subsubsection{Reflectivity}
\begin{figure}[!h]
	\centering
	\includegraphics[width=0.2\textwidth]{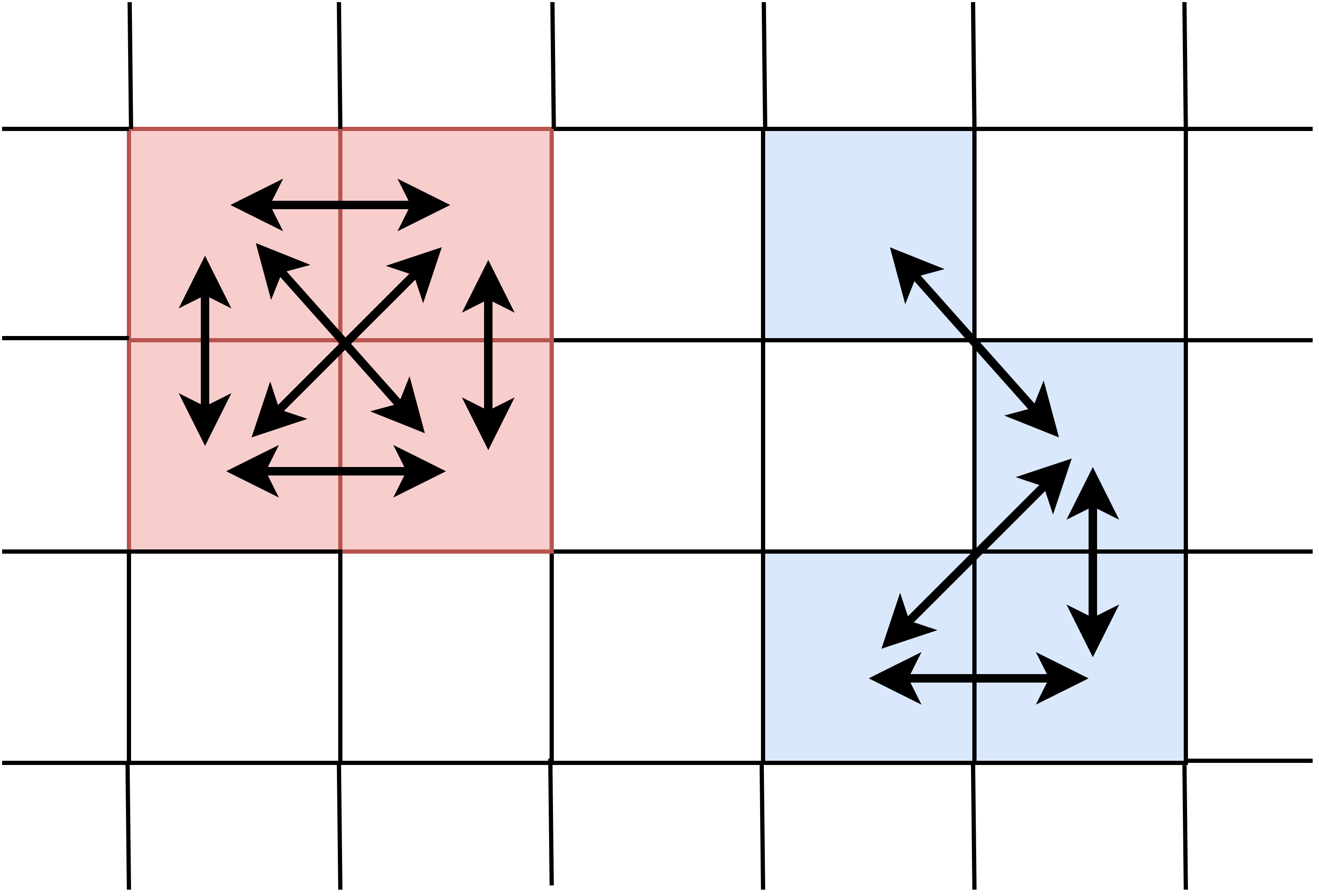}
	\caption{Connectivity at an inter-pixel level. Two different surfaces are denoted by the colours red and blue, where each square represents one pixel.  Pixels without points are represented by white squares. For simplicity, in this example all points are considered to be present at the same depth. Note that each pixel can be connected with at most 8 neighbours.} 
	\label{FIG:connected_pixels}
\end{figure}

The spectral signatures are related to the materials of the surfaces \cite{altmann2017robust}. Neighbouring points corresponding to a surface composed of a specific material show similar spectral signatures. This prior information is added to the model using a Gaussian Markov random field distribution, where the neighbours are defined by the connected-surface structure of the point process prior. First, in order to avoid the positivity constraint on the intensity $r_{n,\ell}$, we use the canonical form \cite{rue2005gaussian,inla,salmon2014poisson},
\begin{equation}
m_{n,\ell} = \log(r_{n,\ell})
\end{equation}
where $m_{n,\ell}\in \mathbb{R}$ denotes the log-intensity of the $n$th point at band $\ell$.
As multispectral devices only acquire dozens of well-separated wavelengths, the spectral measurements within a pixel do not show significant correlation. Hence, although potential correlations between wavelengths could be modelled, we choose here to neglect this correlation to keep the estimation strategy tractable. As a consequence, we consider the following reflectivity prior model 
\begin{equation}
p(\mat{\Phi}_r|\sigma^2,\beta) = \prod_{\ell=1}^{L} p(\myvec{m}_{\ell}|\mat{\Phi}_c,\sigma^2,\beta).
\end{equation}
Spatial correlations between log-intensity values in neighbouring pixels are defined according to the distribution 
\begin{equation} \label{EQ:a_joint}
\myvec{m}_{\ell}|\sigma^2,\beta, \mat{\Phi}_c \sim \mathcal{N}(\myvec{0},\sigma^2\mat{P}^{-1})
\end{equation}
where $\sigma^2$ and $\beta$ are hyperparameters controlling the level of smoothness. The precision matrix $\mat{P}$ is very sparse due to the Markovian structure, being defined by
\begin{equation} \label{EQ:P}
[\mat{P}]_{n,n'}=
\begin{cases}
\beta + \sum_{\tilde{n}\in\mathcal{M}_{pp}(\myvec{c}_n)} \frac{1}{d(\myvec{c}_n;\myvec{c}_{\tilde{n}} )} & \text{if $n=n'$} \\
-\frac{1}{d(\myvec{c}_n;\myvec{c}_{n'} )} & \text{if $\myvec{c}_n\in\mathcal{M}_{pp}(\myvec{c}_{n'})$} \\
0 & \text{otherwise}
\end{cases}
\end{equation} 
where $\mathcal{M}_{pp}(\myvec{c}_{n})$ is the set of neighbours of point $\myvec{c}_{n}$, which is obtained using the connected-surface structure illustrated in \Cref{FIG:connected_pixels}, and $d(\myvec{c}_n;\myvec{c}_{\tilde{n}})$ is the Euclidean distance between two points, normalized according to the camera parameters of the scene to have a physical meaning \cite{hartley2003multiple}.

\subsubsection{Background levels}\label{SUBSEC:bkg_levels}
\begin{figure}[!h]
	\centering
	\includegraphics[width=.48\textwidth]{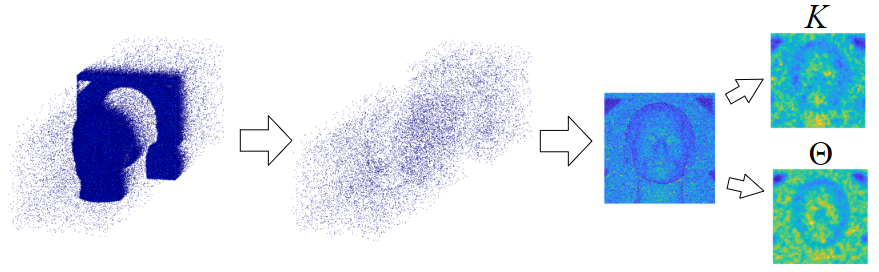}
	\caption{Computation of the hyperparameters for the priors of the background levels. First, the photons due to the signal are removed from the dataset using a coarse approximation of the point cloud. Secondly, the remaining photons are integrated per pixel, giving a noisy background image. Finally, this image is used to estimate uncertainty about the background levels, computing the gamma hyperparameters $\mat{K}$ and $\mat{\Theta}$.}
	\label{FIG:bkg_MSL}
\end{figure}

Background detections are due to detector dark counts and ambient illumination, as explained in \cite{shin2016computational,shin2015photon}. If the transceiver system is mono-static\footnote{In mono-static Lidar systems, the laser and detector are coaxial, whereas in bi-static systems, the source and detector do not share the same axis.} \cite{McCarthy:13}, the set of background levels can be interpreted as a multispectral passive image of the scene, as background detections generally come from ambient illumination reflected onto the target and collected by the single-photon detector. In this case, the background levels are strongly spatially correlated within each wavelength. However, in bi-static systems \cite{shin2016computational}, the transmit and receive channels do not share the same objective lens aperture, yielding weakly or uncorrelated background detections. Although not showing a strong spatial correlation in this second case, all the background levels have similar values, which also serves as prior information. 
\paragraph{Independent prior distributions} 
In order to  simultaneously model potential spatial correlation and ensure the positivity of the background levels, ManiPoP uses a 2D gamma Markov random field, which was introduced by Dikmen et al. in \cite{Dikmen2010}. However, this prior is not well suited for MSL data as it introduces an undesired penalization for large background levels, whose negative effects are amplified when considering multiple bands (see Appendix \ref{APP:gmrf} for details). Other alternatives such as Gaussian Markov random fields \cite{rue2005gaussian} cannot be sampled directly in closed form, requiring proposals with a rejection step, whose mixing and convergence scale badly with the dimension of the spectral cube, as shown in \cite{tachella2018bay}. 
To alleviate these problems, we assign independent gamma priors, i.e.,
\begin{equation}
\left\{
\begin{aligned}
p(\mat{B}|\mat{K},\mat{\Theta}) &= \prod_{i=1}^{N_r} \prod_{j=1}^{N_c}\prod_{\ell=1}^{L} p(
b_{i,j,\ell}|k_{i,j,\ell},\theta_{i,j,\ell})\\
b_{i,j,\ell}|k_{i,j,\ell},\theta_{i,j\ell} &\sim \mathcal{G}(k_{i,j,\ell},\theta_{i,j\ell})
\end{aligned}
\right.
\label{EQ:independent_gamma}
\end{equation}
where $[\mat{\Theta}]_{i,j,\ell}=\theta_{i,j,\ell}$ and $[\mat{K}]_{i,j,\ell}=k_{i,j,\ell}$ are the shape and scale hyperparameters of the gamma distributions. Despite using independent priors, we can capture the spatial correlation by setting the hyperparameters $(\mat{K},\mat{\Theta})$ appropriately. More precisely, in a similar fashion to variational Bayes \cite{beal2003variational} or expectation propagation \cite{minka2001expectation} methods, in order to simplify the estimation of $\mat{B}$, we specify \eqref{EQ:independent_gamma} such that $p(\mat{B}|\mat{K},\mat{\Theta})$ is similar to another distribution $q(\mat{B}) = \prod_{\ell=1}^{L} q_{\ell}(\myvec{b}_{\ell})$ which explicitly correlates the background levels in neighbouring pixels and assumes mutual independence between spectral bands. Here, we use as a similarity criterion the Kullback-Leibler divergence
\begin{equation}
\label{EQ:KL_div}
(\mat{K},\mat{\Theta}) = \argmin_{\mat{K},\mat{\Theta}} \textrm{KL}(q||p). 
\end{equation}
As discussed in Appendix \ref{APP:KL_div}, solving \eqref{EQ:KL_div}  can be achieved by computing expectations with respect to $q(\mat{B})$.
\paragraph{Empirical Bayes approach}
To ensure that the prior model \eqref{EQ:independent_gamma} is informative, a suitable distribution $q(\mat{B})$ should be chosen. Assuming that we have a coarse estimate of the point cloud (this information will be obtained using the multiresolution approach detailed in \Cref{SUBSEC:full algo}), we build the distribution $q(\mat{B})$ following an empirical Bayes approach, as illustrated in \Cref{FIG:bkg_MSL}. First, one can discard almost all the signal photons in the dataset by removing the photons detected in the compact support of $h_{\ell}(t)$ around each point (see \Cref{FIG:bkg_MSL}, central subplot). The number of bins that is not excluded in each pixel is referred to as $v_{i,j,\ell}$. Secondly, we integrate the remaining photons of each pixel, obtaining a coarse estimate of the per-pixel background photon levels, denoted by $\bar{z}_{i,j,\ell,b}$.
We then define $q_\ell(\myvec{b}_{\ell})\propto p(\mat{\bar{z}}_\ell|\myvec{b}_{\ell})p(\mat{b}_{\ell}|\alpha_{B})$ with  
\begin{equation}
\left\{
\begin{aligned}
\bar{z}_{i,j,\ell,b}|b_{i,j,\ell} &\sim \mathcal{P}(g_{i,j,\ell}v_{i,j,\ell}b_{i,j,\ell})\\
b_{i,j,\ell} &= \exp (\tilde{b}_{i,j,\ell} + \mu_\ell) \\
\myvec{\tilde{b}}_{\ell}|\alpha_{B} &\sim \mathcal{N} (\mat{0},\alpha_{B}\mat{D}^{-1})
\label{EQ:poisson-gaussian}
\end{aligned}
\right.
\end{equation}
where $\mat{b}_{\ell}$ a vectorized image of background levels at wavelength $\ell$, $\mat{D}\in\mathbb{R}^{NrNc\times NrNc}$ a positive semidefinite matrix, $\alpha_{B}$ is a fixed hyperparameter controlling the degree of smoothness. In mono-static systems, a two-dimensional Laplacian filter is chosen for $\mat{D}$ to promote spatial correlation \cite{rue2005gaussian}, whereas in bi-static systems, $\mat{D}$ is replaced by the identity matrix, only penalizing large background levels. $\mu_\ell$ is a translation parameter centring $\tilde{b}_{i,j,\ell}$ in the linear part of the exponential function and is defined as $\mu_\ell=\log(\frac{1}{N_rN_c}\sum_{i}^{N_r}\sum_{j}^{N_c}\frac{\bar{z}_{i,j,\ell,b}g_{i,j,\ell}}{v_{i,j,\ell}})$. As mentioned above, solving  \eqref{EQ:KL_div} requires the computation of expectations with respect to $q(\mat{B})$ which are unfortunately not available in closed form. Instead of using additional MCMC sampling to find numerical approximations (detailed in Appendix \ref{APP:KL_div}), obtaining samples from \eqref{EQ:poisson-gaussian} is more attractive both in terms of convergence and computational complexity than using the original full datacube which includes mixtures of background and signal photons. Indeed, \eqref{EQ:poisson-gaussian} simply involves integrated photon counts (over the range dimension). Moreover, given the independence property of $q(\mat{B})$ among spectral bands, all the bands can be processed independently in parallel \textcolor{black}{when sampling $\mat{B}$ (see \Cref{SUBSEC:bkg sampling})}.

\subsection{Posterior distribution}
\begin{figure}[!h]
	\centering
	\includegraphics[width=0.38\textwidth]{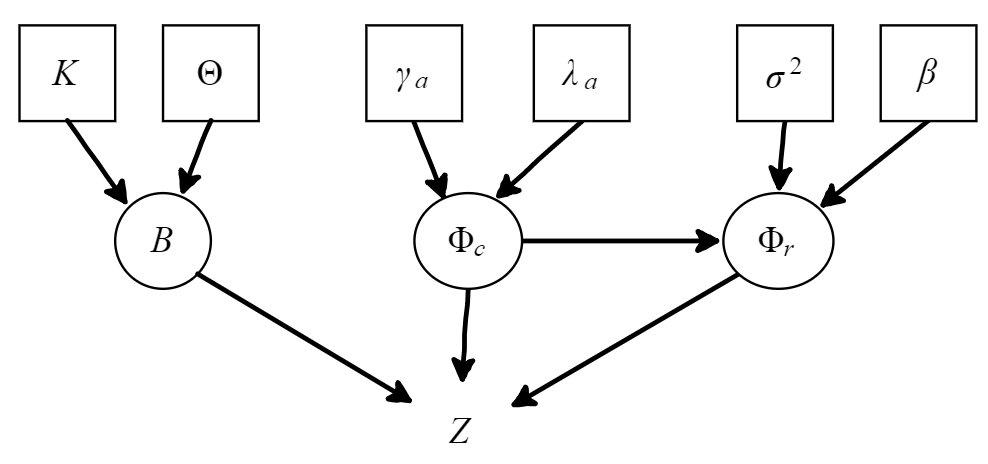}
	\caption{Directed acyclic graph (DAG) of the proposed hierarchical Bayesian model. The variables inside squares are fixed, whereas the variables inside circles are estimated.}
	\label{FIG:dag}
	\vspace{-0.3cm}
\end{figure}

Following Bayes theorem, the joint posterior distribution of the model parameters is given by
\begin{multline}
\label{EQ:posterior}
p(\mat{\Phi}_c,\mat{\Phi}_r,\mat{B}|\mat{Z},\Psi) \propto p(\mat{Z}|\mat{\Phi}_c,\mat{\Phi}_r,\mat{B})  p(\mat{\Phi}_r|\mat{\Phi}_c,\sigma^2,\beta) \times \\  f_1(\mat{\Phi}_c|\gamma_{a},\lambda_{a}) f_2(\mat{\Phi}_c|\gamma_{st})\pi(\mat{\Phi}_c)p(\mat{B}|\mat{K},\mat{\Theta})
\end{multline}
where $\Psi$ denotes the set of hyperparameters  $\Psi=\{\gamma_{a},\lambda_{a},\gamma_{s},\sigma^2,\beta,\mat{K},\mat{\Theta}\}$ and $\pi(\cdot)$ is the Poisson point process reference measure. \Cref{FIG:dag} shows the directed acyclic graph associated with the proposed hierarchical Bayesian model.

\section{Inference} \label{SEC:RJ-MCMC}
\textcolor{black}{In this work, we compute the same posterior statistics as in \cite{tachella2018manipop}: the point cloud positions and spectral signatures are estimated using the maximum-a-posteriori (MAP) estimator}
\begin{equation}
\mat{\hat{\Phi}}=\argmax_{\mat{\Phi}} p(\mat{\Phi},\mat{B}|\mat{Z},\Psi)
\end{equation}
\textcolor{black}{and the minimum mean squared error estimator is considered for $\mat{B}$}
\begin{equation}
\mat{\hat{B}}=\mathbb{E}\{\mat{B}|\mat{Z},\Psi\}.
\end{equation}
As this expectation cannot be derived analytically, we propose to investigate Markov Chain Monte Carlo (MCMC) simulation methods. \textcolor{black}{As in \cite{tachella2018manipop},} we use a reversible jump MCMC algorithm that can handle the varying dimension nature of the spatial point process. This sampler generates $N_{m}$ samples of $\mat{\Phi}$ and $\mat{B}$ from the posterior distribution \eqref{EQ:posterior} denoted as
\begin{equation}
\{\mat{\Phi}^{(s)} ,\mat{B}^{(s)}   \quad \forall s=0,1,\dots, N_{m}-1 \}.
\end{equation}
These samples are then used to approximate the statistics of interest, i.e.,
\begin{align*}
\mat{\hat{\Phi}} &\approx \argmax_{s=0,...,N_m-1} p(\mat{\Phi}^{(s)} , \mat{B}^{(s)} |\mat{Z},\Psi) \\
\mat{\hat{B}} &\approx \frac{1}{N_{m}-N_{\textrm{bi}}}\sum_{s=N_{\textrm{bi}}+1}^{N_{m}}\mat{B}^{(s)}
\end{align*}
where $N_{\textrm{bi}}$ is the number of burn-in iterations.

\subsection{Reversible jump MCMC moves}
While other stochastic simulation algorithms for varying dimensions can be used \cite{brooks2011handbook}, we choose an RJ-MCMC sampler, as it allows us to build proposals tailored for the MSL reconstruction problem. RJ-MCMC can be interpreted as a natural extension of the Metropolis-Hastings algorithm for problems with an a priori unknown dimensionality. Given the actual state of the chain $\myvec{\theta} =\{\mat{\Phi} , \mat{B} \}$ of model order $N_{\Phi }$, a random vector of auxiliary variables $\myvec{u}$ is generated to create a new state  $\myvec{\theta}'=\{\mat{\Phi}',\mat{B'}\}$ of model order $N_{\Phi'}$, according to an appropriate deterministic function $\myvec{\theta}'=g(\myvec{\theta} ,\myvec{u})$. To ensure reversibility, an inverse mapping with auxiliary random variables $\myvec{u'}$ has to exist such that $\myvec{\theta} =g^{-1}(\myvec{\theta}',\myvec{u'})$. The move $\myvec{\theta} \rightarrow\myvec{\theta}'$ is accepted or rejected with probability $\rho=\min\{1,r\left(\myvec{\theta} ,\myvec{\theta}'\right)\}$, where $r(\cdot,\cdot)$ is defined as
\begin{equation}
\label{EQ:RevJump}
r\left(\myvec{\theta},\myvec{\theta'}\right) = \frac{p(\myvec{\theta'}|\mat{Z},\Psi)K(\myvec{\theta}|\myvec{\theta'})p(\myvec{u'})}{p(\myvec{\theta}|\mat{Z},\Psi)K(\myvec{\theta'}|\myvec{\theta})p(\myvec{u})}\left|\frac{\partial g(\myvec{\theta},\myvec{u})}{\partial(\myvec{\theta},\myvec{u})}\right|
\end{equation}
and $K(\myvec{\theta'}|\myvec{\theta})$ is the probability of proposing the move $\myvec{\theta}\rightarrow\myvec{\theta'}$, $p(\myvec{u})$ is the probability distribution of the random vector $\myvec{u}$, and $\left|\frac{\partial g(\myvec{\theta},\myvec{u})}{\partial(\myvec{\theta},\myvec{u})}\right|$ is the Jacobian of the mapping $g(\cdot)$. The RJ-MCMC algorithm performs birth/death, dilation/erosion, spatial and mark shifts, and split/merge moves with probabilities $p_{\textrm{birth}}$, $1-p_{\textrm{birth}}$, $p_{\textrm{dilation}}$, $1-p_{\textrm{dilation}}$, $p_{\textrm{shift}}$, $p_{\textrm{mark}}$, $p_{\textrm{split}}$ and $1-p_{\textrm{split}}$ respectively. Due to the Markovian nature of the prior distributions, all the proposed moves are \emph{local}, having a complexity proportional to the size of the neighbourhood. These moves are detailed in the following subsections. For ease of presentation, we summarize the main aspects of each move, inviting the reader to consult Appendix \ref{APP:moves} for more details.

\paragraph{Birth and death moves}
The birth move proposes a new point $\myvec{\theta'}=(\mat{c}_{N_\Phi+1},\mat{m}_{N_\Phi+1})$ uniformly at random in the 3D cube. The spectral signature of the new point is proposed by extracting a fraction $(1-u_{\ell})$ from the current value of the background level $b_{i,j,\ell}$ according to the signal-to-background-ratio (SBR) \cite{Rapp2017AFP} $w_{\ell}$, that is for each wavelength $\ell$

\begin{equation}
\left\{
\begin{aligned}
\textcolor{black}{u}_{\ell} & \sim \mathcal{U}(0,1),  b_{i,j,\ell}' =  \textcolor{black}{u}_{\ell} b_{i,j,\ell}\\
e^{m_{N_\Phi+1,\ell}} &= w_{\ell} (1-\textcolor{black}{u}_{\ell}) b_{i,j,\ell}\frac{T}{\sum_{t=1}^{T}h_{\ell}(t)} 
\end{aligned}
\right.,
\label{EQ:birth_update}
\end{equation}
where $\mathcal{U}(0,1)$ denotes the uniform distribution on the interval $(0,1)$.
The death move proposes the removal of a point. In contrast to the birth move, we modify the background level according to
\begin{equation}
b_{i,j,\ell}' = b_{i,j,\ell} + e^{m_{N_\Phi+1,\ell}}\frac{\sum_{t=1}^{T}h_{\ell}(t)}{w_{\ell} T} \quad \forall \ell=1,\dots,L.
\label{EQ:death_move}
\end{equation}

\paragraph{Dilation and erosion moves}
Birth moves have low acceptance ratio, as the probability of randomly proposing a point within or close to the surfaces of interest is very low. However, this problem can be overcome by using the current estimation of the surface to propose in regions of high probability.
The dilation move proposes a point inside the neighbourhood of an existing surface with uniform probability across all possible neighbouring positions where a point can be added. \textcolor{black}{Contrary to \cite{tachella2018manipop}, where the intensity samples are generated according to the prior distribution, }the spectral signature is sampled in the same way as the birth move \eqref{EQ:birth_update}.  The complementary move (named erosion), proposes to remove a point $\myvec{c}_{n}$ with one or more neighbours. In this case, the background is updated in the same way as the death move.
\paragraph{Mark and shift moves}
\textcolor{black}{As in ManiPoP}, the mark move updates the log-intensity of a randomly chosen point $\myvec{c}_n$. Each wavelength is updated independently using a Gaussian proposal with variance $\delta_{m}$ as a proposal (also known as Metropolis Gaussian random walk), that is 
\begin{equation}\label{EQ:mark_prop}
m'_{n,\ell} \sim \mathcal{N}\left(m_{n,\ell} ,\delta_{m}\right) \quad \forall \ell=1,\dots,L.
\end{equation}
Similarly, the shift move updates the position of a uniformly chosen point using a Gaussian proposal with variance $\delta_{t}$
\begin{equation}\label{EQ:shift_prop}
t'_{n} \sim \mathcal{N}\left(t_n ,\delta_{t}\right)
\end{equation}
The values of $\delta_{m}$ and $\delta_{t}$ are adjusted by cross-validation\footnote{\textcolor{black}{Intensities are normalized to belong to a fixed interval across datasets. Hence, we can fix the variance of the proposal to achieve similar acceptance ratios.}} to yield an acceptance ratio close to $41\%$ for each move, which is the optimal value \textcolor{black}{for a one dimensional Metropolis random walk}, as explained in \cite[Chapter~4]{brooks2011handbook}. 
\paragraph{Split and merge moves}
Some pixels might present two points with overlapping impulse responses in depth. In such cases, a death move followed by two birth moves would happen with very low probability. Hence, as in ManiPoP, we propose a split move, which randomly picks a point $(\myvec{c}_n,\myvec{m}_n)$ and proposes two new points, $(\myvec{c}'_{k_1},\myvec{m}'_{k_1})$ and $(\myvec{c}'_{k_2},\myvec{m}'_{k_2})$. The log-intensity is proposed for each wavelength following the mapping
\begin{equation}
\left\{
\begin{aligned}
u_{\ell} &\sim \mathcal{B}(\eta,\eta) \\
m'_{k_1,\ell} &= m_{n,\ell} + \log(u_{\ell}) \\
m'_{k_2,\ell} &= m_{n,\ell} + \log(1-u_{\ell})  
\end{aligned}
\right.
\end{equation}
where $\mathcal{B}(\cdot)$ denotes the beta distribution and $\eta$ is a fixed parameter. The new positions are determined according to 
\begin{equation}
\left\{
\begin{aligned}
s_{\ell} &\sim \mathcal{B}\textrm{e}(0.5) \\
\Delta &\sim \mathcal{U}(d_{\min},L_h) \\
t'_{k_1} &= t_{n} +(-1)^{s_{\ell}} \Delta \frac{\sum_{\ell=1}^{L}(1-u_{\ell})e^{m_{n,\ell}}}{\sum_{\ell=1}^{L}e^{m_{n,\ell}} } \\
t'_{k_2} &= t_{n} - (-1)^{s_{\ell}} \Delta \frac{\sum_{\ell=1}^{L} u_{\ell} e^{m_{n,\ell}}} {\sum_{\ell=1}^{L}e^{m_{n,\ell}}}
\end{aligned}
\right.
\end{equation}
where $\mathcal{B}$e$(\cdot)$ denotes the Bernoulli distribution, $L_h$ is the length in bins of the instrumental response at the wavelength with longest impulse response.
The complementary move, named merge move, is performed by randomly choosing two points $\myvec{c}_{k_1}$ and $\myvec{c}_{k_2}$ inside the same pixel ($x_{k_1}=x_{k_2}$ and $y_{k_1}=y_{k_2}$) that satisfy the condition 
\begin{equation}
\label{EQ: merge_condition}
d_{\min}<\left|t_{k_1}-t_{k_2}\right| \le L_h \quad \forall \ell=1,\dots,L
\end{equation}
The merged point $(\myvec{c'}_n,\mat{r}'_n)$ is obtained by the inverse mapping
\begin{equation*}
\left\{
\begin{aligned}
\label{EQ:merge_mapping}
&e^{m'_{n,\ell}} = e^{m_{k_1,\ell}} + e^{m_{k_2,\ell}}  \quad \forall \ell=1,\dots,L \\
&t'_{n} = t_{k_1} \frac{\sum_{\ell=1}^{L}e^{m_{k_1,\ell}}}{\sum_{\ell=1}^{L}e^{m_{k_1,\ell}} + e^{m_{k_2,\ell}}} +t_{k_2} \frac{\sum_{\ell=1}^{L}e^{m_{k_2,\ell}}}{\sum_{\ell=1}^{L}e^{m_{k_1,\ell}} + e^{m_{k_2,\ell}}}
\end{aligned}
\right.
\end{equation*}
which preserves the total pixel intensity and weights the spatial shift of each peak according to the sum of the intensity values.

\subsubsection{Sampling the background}\label{SUBSEC:bkg sampling}
In order to exploit the conjugacy between the Poisson likelihood and gamma priors for the background levels, we use a data augmentation scheme as in \cite{zhou2012beta}, which classifies each photon-detection according to the source (target(s) or background), i.e.,
\begin{align*}
z_{i,j,t,\ell} =& \sum_{\substack{n : (x_n,y_n)=(i,j) }} \tilde{z}_{i,j,t,\ell,n} + \tilde{z}_{i,j,t,\ell,b} \\
\tilde{z}_{i,j,t,\ell,b} \sim&  \mathcal{P}(g_{i,j,\ell} b_{i,j,\ell}) \\
\tilde{z}_{i,j,t,\ell,n} \sim&  \mathcal{P}(g_{i,j,\ell} r_{n,\ell} h(t-t_n))
\end{align*}
where $\tilde{z}_{i,j,t,\ell,n}$ are the photons at bin $t$ associated with the $n$th surface and $\tilde{z}_{i,j,t,\ell,b}$ are the ones associated with the background. In the augmented space defined by $(\tilde{z}_{i,j,t,\ell,n},\tilde{z}_{i,j,t,\ell,b})$, the background levels are sampled as follows
\begin{equation}
\label{EQ:B_scheme}
\left\{
\begin{aligned}
\tilde{z}_{i,j,t,\ell,b} &\sim \mathcal{B}\left(z_{i,j,l,t},\frac{b_{i,j,\ell}}{\sum_{\substack{n:(x_n,y_n)=(i,j)}}r_{n,\ell}h(t-t_{n})+b_{i,j,\ell}}\right) 
\\ b_{i,j,\ell} & \sim \mathcal{G} \left(r_{i,j,\ell}+\sum_{t=1}^{T}\tilde{z}_{i,j,t,\ell,b},\frac{\theta_{i,j,\ell}}{g_{i,j,\ell}T\theta_{i,j,\ell}+1}\right)
\end{aligned}
\right.
\end{equation}
where $\mathcal{B}(\cdot)$ denotes the Binomial distribution. The transition kernel defined by \eqref{EQ:B_scheme} produces samples of $b_{i,j,\ell}$ distributed according to the marginal posterior distribution of $\mat{B}$. In practice, we observed that only one iteration of this kernel is sufficient.

\subsection{Full algorithm} \label{SUBSEC:full algo}
We adopt a multi-resolution approach to speed up the convergence of the RJ-MCMC algorithm, in a fashion similar to \cite{tachella2018manipop}. The dataset is downsampled by integrating photon detections in patches of $N_{\textrm{bin}}\times N_{\textrm{bin}}$ pixels. Hence, the number of pixels is reduced by a factor of $N_{\textrm{bin}}^2$, meaning less points and background levels to infer with $N_{\textrm{bin}}^2$ times more photons per pixel. The estimated point cloud at the coarse scale is upsampled using a simple nearest neighbour algorithm and used as initialization for the next (finer) scale. In all our experiments we repeat the process for $K=3$ scales. The background hyperpriors $\mat{K}$ and $\mat{\Theta}$ are initialized with non-informative values, i.e., $k_{i,j,\ell}=0.01$ and $\theta_{i,j,\ell}=100$ for all pixels $(i,j)$ and wavelengths $\ell$. In finer scales, these hyperparameters are computed using the algorithm detailed in \Cref{SUBSEC:bkg_levels}. The multi-resolution approach is finally summarized in \Cref{ALG: MR_approach}.
\begin{algorithm}
	\caption{Multiresolution MuSaPoP}
	\label{ALG: MR_approach}
	\begin{algorithmic}[h]
		\STATE \textbf{Input:} MSL waveforms $\mat{Z}$, hyperparameters $\Psi$,  window size $ N_{\textrm{bin}}$ and number of scales $K$.
		\STATE \textbf{Initialization:}
		\STATE $\mat{\Phi}_1^{(0)} \gets \emptyset $
		\STATE $ \mat{B}_1^{(0)}  \gets$  sample from \eqref{EQ:B_scheme}
		\STATE $(\mat{K}_1,\mat{\Theta}_1) \gets$  non-informative hyperparameter values
		\STATE \textbf{Main loop:}
		\FOR{$k = 1,\dots,K$}
		\IF {$k>1$}
		\STATE $(\mat{\Phi}_k^{(0)},\mat{B}_k^{(0)}) \gets$ upsample$(\hat{\mat{\Phi}}_{k-1},\mat{\hat{B}}_{k-1})$
		\STATE Compute hyperparameters $(\mat{K}_k,\mat{\Theta}_k)$ and SBR using \Cref{SUBSEC:bkg_levels}
		\ENDIF
		\STATE $(\mat{\hat{\Phi}}_k,\mat{\hat{B}}_k) \gets$MuSaPoP$(\mat{Z}_k,(\mat{\Phi}_k^{(0)},\mat{B}_k^{(0)}),\Psi_k,\textrm{SBR})$ 
		\ENDFOR
		\STATE \textbf{Output:}  $(\hat{\mat{\Phi}}_K,\mat{\hat{B}}_K)$
	\end{algorithmic}
\end{algorithm}

\section{Subsampling strategy}\label{SEC:CS}
\begin{figure}[!h]
	\centering
	\subfloat[]{
		\includegraphics[width=.4\linewidth]{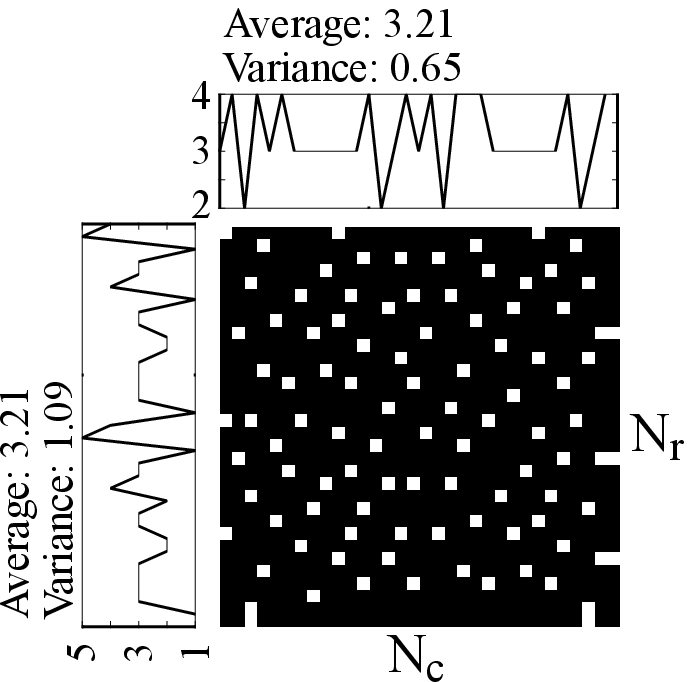}
	}
	\subfloat[]{
		\centering
		\includegraphics[width=.4\linewidth]{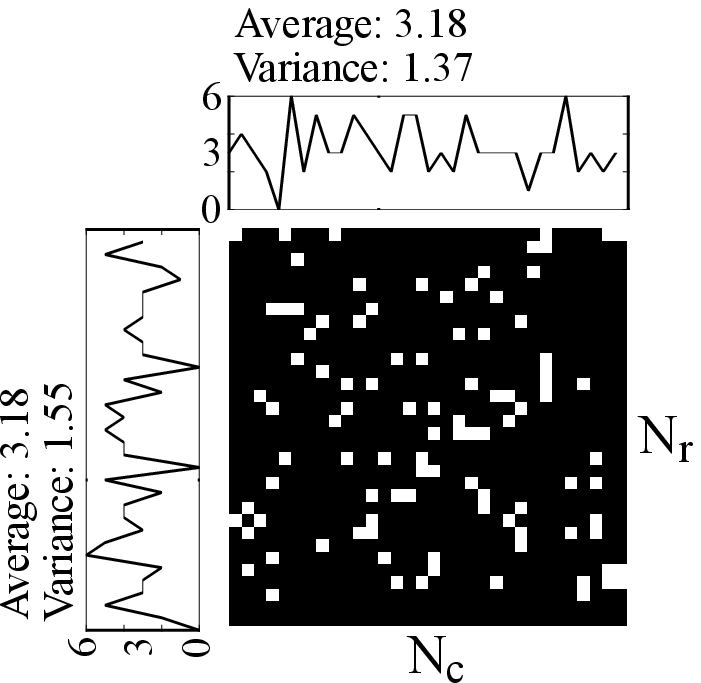} 
	}
	\caption{Subsampling strategies for a Lidar cube with $L=8$ wavelengths, $N_r=N_c=32$ pixels and total compression of $1/L$, i.e., one observed band per pixel. The sampled pixels at the first wavelength are shown in white. A completely random strategy \cite{altmann2017undersample} is shown \textcolor{black}{in (b), whereas the one proposed here is shown in (a).}}
	\label{FIG:sampling_codes}
\end{figure}

Despite not being able to design compressive measurements along the depth axis, we can still reduce the number of measurements in the two spatial (horizontal and vertical) dimensions and in the spectral dimension \cite{altmann2017undersample}. Given the point positions, recovering their reflectivity profile reduces to a multispectral  image restoration problem using measured data corrupted by Poisson noise. While many compressive sensing strategies have been proposed for measurements under this noise assumption \cite{Raginsky2010compressed,Raginsky2010expander,Li2018Minimax}, MSL datasets have an additional limitation if multiple surfaces per pixel are considered: photon-detections belonging to different wavelengths cannot be integrated into a single histogram, as the reconstruction problem \textcolor{black}{generally} becomes highly non-convex, preventing practical reconstruction algorithms\footnote{As mentioned in the introduction, the system presented in \cite{Ren:18} considers the integration of photons belonging to different histograms, but is limited to one surface per pixel.}. Indeed, summing histograms associated with different wavelengths and including multiple peaks generates histograms with even more peaks (possibly overlapping), which makes the 3D reconstruction and the reflectivity estimation more difficult. As a consequence, we only consider subsampling of depth histograms, which incorporates all of the practical sampling limitations. Following the formulation of the observation model \eqref{EQ:full_likelihood}, the subsampling strategy consists of choosing the binary coefficients $\mat{G}$ for a given compression level $W/L$, with $W$ the average number of observed band per pixel. Several subsampling strategies have been proposed for different applications, such as halftoning and stippling \cite{deussen2000floating}, rendering, compressive spectral imaging \cite{galvis2019shifting,galvis2017coded,leon2019temporal}, compressive computed tomography \cite{mojica2017high,choi2009coded},  geometry processing \cite{hinojosa2018coded}, amongst others \cite{surazhsky2003isotropic,deussen1998realistic,liang2015poisson}. These approaches exploit the sampling geometry of the sensing devices to design a set of criteria. Similarly to coded aperture snapshot spectral imaging systems, the distribution of reflectivity profiles of 3D surfaces in real natural scenes suggests uniform sampling in the row, column and spectral axes. Following the design in Correa et al. \cite{correa2016spatiotemporal}, the coefficients are chosen according to the spatiotemporal characteristics of blue noise, which distributes the measurements in spectral and spatial dimensions as homogeneously as possible \cite{Lau2003blue}. The binary cube $\mat{G}$ is obtained by minimizing the variance of (weighted) measurements per local neighbourhood, i.e.,
\begin{align*}
 \argmin_{\mat{G}} &  \quad \textrm{VAR}\left\{\sum_{\ell=1}^L \sum_{(i',j')\in\mathcal{M}^{(i,j)}_{ss}}  w_{i',j'}g_{i',j',\ell}  \right\} \\
\text{subject to} \quad & \sum_{\ell=1}^{L} g_{i,j,\ell} =W \quad \forall (i,j)
\end{align*}
where $\textrm{VAR}\{\cdot\}$ denotes the variance operator, $\mathcal{M}^{(i,j)}_{ss}$ denotes the set of pixels in a local neighbourhood of $(i,j)$ and $w_{i',j'}$ are the weights. The minimization is simplified by dividing the data in slices of $L/W$ contiguous bands and running the algorithm introduced in \cite{correa2016spatiotemporal} per slice. As shown in \Cref{FIG:sampling_codes}, the proposed strategy distributes the measurements uniformly in space, while other random strategies \cite{altmann2017undersample} tend to exhibit clusters, leaving some regions without measurements.

\section{Experiments}\label{SEC:Exper}
To illustrate the efficacy of the proposed method, the new reconstruction algorithm is compared to other alternatives \textcolor{black}{(based on the work conducted in \cite{wallace2014msl})} using a synthetic dataset. Subsequently, the new subsampling scheme is compared with other random subsampling choices for a real MSL dataset. In all the experiments, the performance was measured using the following summary statistics:
\begin{itemize}
	\item True detections $F_{\textrm{true}}(\tau)$: Probability of true detection as a function of the distance $\tau$, considering an estimated point as a true detection if there is another point in the ground truth point cloud in the same pixel ($x^{\textrm{true}}_n=x^{\textrm{est}}_{n'}$ and $y^{\textrm{true}}_n=y^{\textrm{est}}_{n'}$) such that $|t^{\textrm{true}}_{n}-t^{\textrm{est}}_{n'}|\le\tau$.
	\item False detections $F_{\textrm{false}}(\tau)$: Number of estimated points that cannot be assigned to a ground truth point at a distance $\tau$. 
	\item Mean intensity absolute error at distance $\tau$ (IAE): Mean across all the points of the intensity absolute error $\sum_{\ell=1}^{L} |r^{\textrm{true}}_{n,\ell}-r^{\textrm{est}}_{n',\ell}|$, normalized with respect to the total number of ground truth points. The ground truth and estimated points are coupled using the probability of detection $F_{\textrm{true}}(\tau)$. Note that if a point was falsely estimated or a ground truth point was not found, then they are considered to have resulted in an error of  $\sum_{\ell=1}^{L} |r_{n,\ell}|$. The comparison is done with normalized intensity values, that is $\sum_{t=1}^T h_{\ell}(t)=1$ for $\ell=1,\dots,L$.
	\item Background normalized mean squared error $\textrm{NMSE}_{\mat{B}}$: Mean of the normalized squared error of the estimated background at each wavelength, i.e., $\frac{1}{L} \sum_{\ell=1}^{L} \frac{\sum_{i=1}^{N_r}\sum_{j=1}^{N_r} (b^{\textrm{true}}_{i,j,\ell}-\hat{b}_{i,j,\ell})^2}{\sum_{i=1}^{N_r}\sum_{j=1}^{N_r} {\left(b^{\textrm{true}}_{i,j,\ell} \right)}^2}$.
\end{itemize}

\begin{table}
	\centering
	\begin{tabular}{| c | c | c |}
		\hline
		Hyperparameter & Coarse scale                            & Fine scale                    \\ \hline
		$\gamma_{a}$  & $e^{2}$                                 & $e^{3}$                       \\ \hline
		$\lambda_{a}$ & $(N_rN_r/N_{\textrm{bin}}^2)^{1.5}$          & $(N_rN_r)^{1.5}$ \\ \hline
		$N_{b}$      & $8N_{\textrm{bin}}\Delta_{p}/\Delta_{\textrm{bin}}$ & $8\Delta_{p}/\Delta_{\textrm{bin}}$  \\ \hline
		$d_{\min}$      & $2N_{b}+1$                            & $2N_{b}+1$                  \\ \hline
		$\sigma^2 $    & $0.6^2$                                 & $0.6^2/N_{\textrm{bin}}$                       \\ \hline
		$\beta $       & $\sigma^2/100$                           & $\sigma^2/100$                 \\ \hline
	\end{tabular}
	\caption{Hyperparameters values}
	\label{TAB:hyperparameters}
	\vspace{-.7cm}
\end{table}

\subsection{Synthetic data}
\begin{figure}[!h]
	\centering
	\includegraphics[width=.5\linewidth]{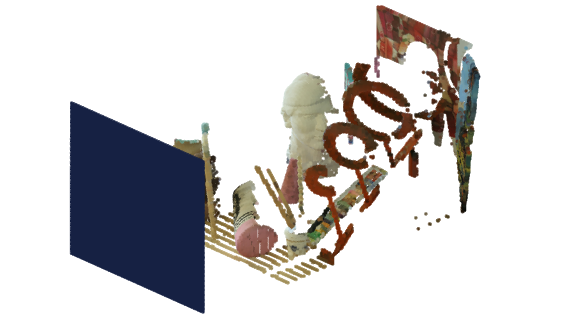}
	\caption{\textcolor{black}{Synthetic ``Art'' scene from Middlebury dataset with an additional semitransparent surface (blue plane).}}
	\label{FIG:synthetic 3D}
	\vspace{-0.1cm}
\end{figure}

\begin{figure}[!h]
	\centering
	\includegraphics[width=1\linewidth]{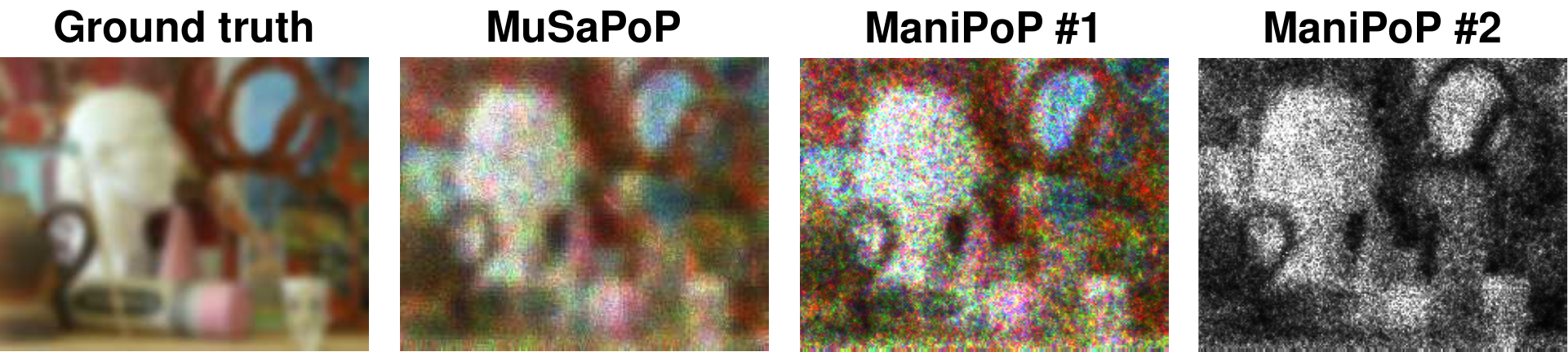}
	\caption{\textcolor{black}{From left to right: Ground truth background levels, estimates obtained by MuSaPoP and the two ManiPoP alternatives. Only the red, blue and green channels were used to generate these images. The proposed method provides smooth estimates due to the empirical prior distribution described in \Cref{SUBSEC:bkg_levels}.} \textcolor{black}{ManiPoP \#2 only estimates one wavelength, which is shown in grayscale.}}
	\label{FIG:synthetic bkg}
	\vspace{-0.3cm}
\end{figure}

We first assessed the performance of the proposed algorithm using a synthetic dataset created from the ``Art'' scene of the Middlebury dataset \cite{middlebury}, as shown in \Cref{FIG:synthetic 3D}. The measurements were obtained by simulating the single-photon multispectral Lidar system of \cite{altmann2017robust}, whose bin width is 0.3 mm. The generated dataset has $N_r=283$ and $N_c=231$ pixels\textcolor{black}{, $T=4500$ histogram bins and $L=4$ wavelengths (red, green, blue and yellow), where only $W=2$ wavelengths out of 4 were sampled per pixel using the coded aperture introduced in \Cref{SEC:CS}. The mean number of photons per wavelength per pixel is 10, where approximately 3.4 photons are due to the background illumination. As in mono-static Lidar systems, the background levels are generated as a passive image of the scene (see \Cref{FIG:synthetic bkg}).}

We compared the proposed method to a two-stage algorithm that first estimates the point positions using ManiPoP \cite{tachella2018manipop} (single-wavelength multiple-return state-of-the-art algorithm) \textcolor{black}{by integrating the photons across wavelengths} and then infers the spectral signatures with fixed point positions, \textcolor{black}{similarly to} the procedure suggested by Wallace et al. \cite{wallace2014msl}. \textcolor{black}{The resulting method, referred to as ManiPoP \#1, is summarized in \Cref{ALG: two-stage}. We also compared with ManiPoP in the strict single-wavelength setting, by choosing the most powerful wavelength and using the same total acquisition time per pixel than in the multispectral case (i.e., a per-pixel acquisition time $W=2$ longer than the one considered for each wavelength in MuSaPoP). This second alternative is referred to as ManiPoP \#2.}
\begin{algorithm}
	\caption{ManiPoP \#1\cite{wallace2014msl,tachella2018manipop}}
	\label{ALG: two-stage}
	\begin{algorithmic}[h]
		\STATE \textbf{Input:} MSL waveforms $\mat{Z}$
		\STATE \textbf{Depth estimation:} \textcolor{black}{Accumulate photons across wavelengths $z'_{i,j,t}=\sum_{\ell}z_{i,j,\ell,t}$ for all pixels $(i,j)$} 
		\STATE $(\mat{\hat{\Phi}},\mat{\hat{B}}) \gets$ManiPoP\textcolor{black}{$(\mat{Z'})$}
		\FOR{$\ell = 1,\dots,L$ and $\ell\ne w$}
		\STATE Update $(\mat{\hat{\Phi}},\mat{\hat{B}})$ using \textcolor{black}{ManiPoP$(\mat{Z}_{\ell})$} in a fixed dimensional setting (only using background and reflectivity moves)
		\ENDFOR
	\end{algorithmic}
\end{algorithm}

\begin{figure}[!h]
	\centering
	\includegraphics[width=.44\textwidth]{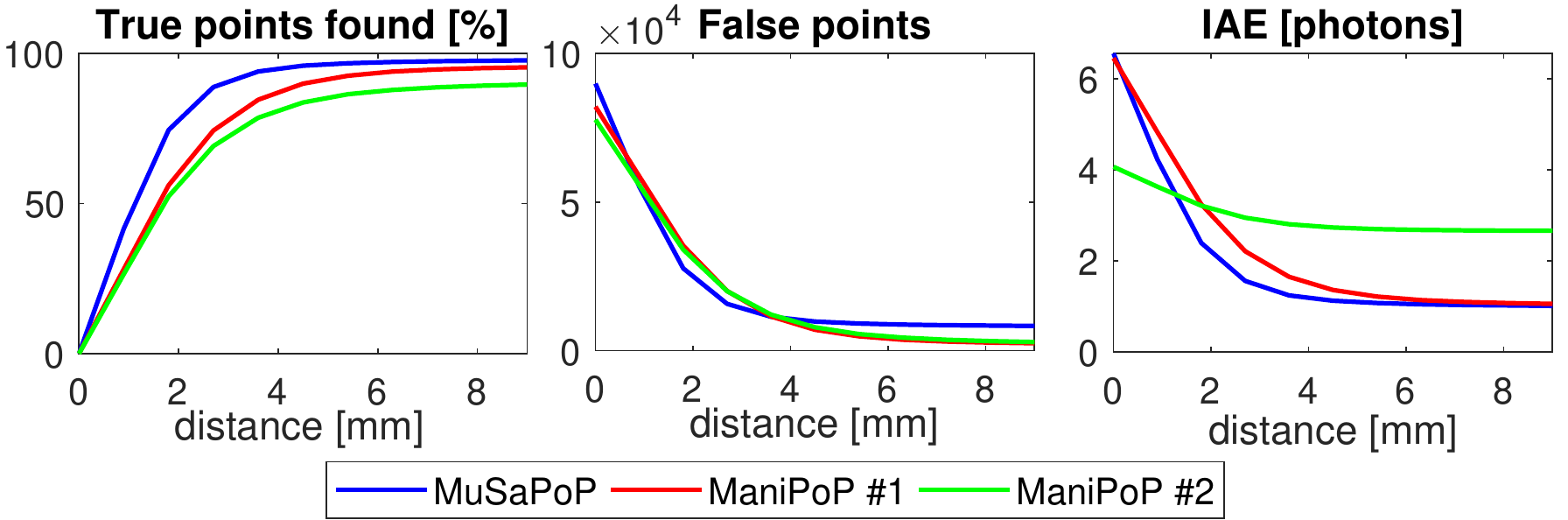}
	\caption{From left to right: $F_{\textrm{true}}(\tau)$, $F_{\textrm{false}}(\tau)$ and IAE$(\tau)$ for the proposed method and the \textcolor{black}{two alternatives.}}
	\label{FIG:synthetic_results}
\end{figure}

\textcolor{black}{\Cref{FIG:synthetic_results} illustrates $F_{\textrm{true}}(\tau)$,  $F_{\textrm{false}}(\tau)$ and IAE for both methods. The proposed algorithm performs better than the other alternatives, as it finds 97.7\% of the true points, whereas  ManiPoP \#1 and \#2 only recover 95.34\% and 89.6\% respectively. ManiPoP \#1 relies on an approximate impulse response $\tilde{h}(t)=\sum_{\ell=1}^{L}h_\ell(t)$, which biases the depth estimates, as the true accumulated response varies across points depending on their spectral signature\footnote{Note that this bias can be arbitrarily large depending on the variations of $h_\ell(t)$ across wavelengths.}. \textcolor{black}{The bias degrades the performance in terms of average depth absolute error (computed for true detections within a distance of 9 mm from the ground truth point). The proposed method obtains an average error of 3.9 mm, whereas the estimates by ManiPoP \#1 present an average error of 5.7 mm.} Despite having double acquisition time for the single-wavelength, ManiPoP \#2 fails to find points corresponding to materials that have very low reflectivity in the blue wavelength (e.g., the red helicoidal structure shown in \Cref{FIG:synthetic 3D}). The proposed MuSaPoP algorithm performs slightly worse in terms of false detections, finding 3 times more false points than the competitor methods. In terms of intensity estimation, MuSaPoP obtains better results, having an asymptotic IAE of 1 photon, whereas the alternatives \#1 and \#2 provide IAE equal to 1.1 and 2.7. The estimated background levels are shown in \Cref{FIG:synthetic bkg}. The proposed method yields a better background NMSE (0.04) than alternatives \#1 (0.14) and \#2 (0.79). The improvement in background estimation over the ManiPoP alternatives can be attributed to the use of an empirical Bayes prior instead of a gamma Markov random field. The total execution time was 811 s for MuSaPoP and 294 and 348 s for alternatives \#1 and \#2.}

\subsection{Real MSL data}
\begin{figure}[!h]
	\centering
	\subfloat[]{
		\includegraphics[width=0.28\linewidth]{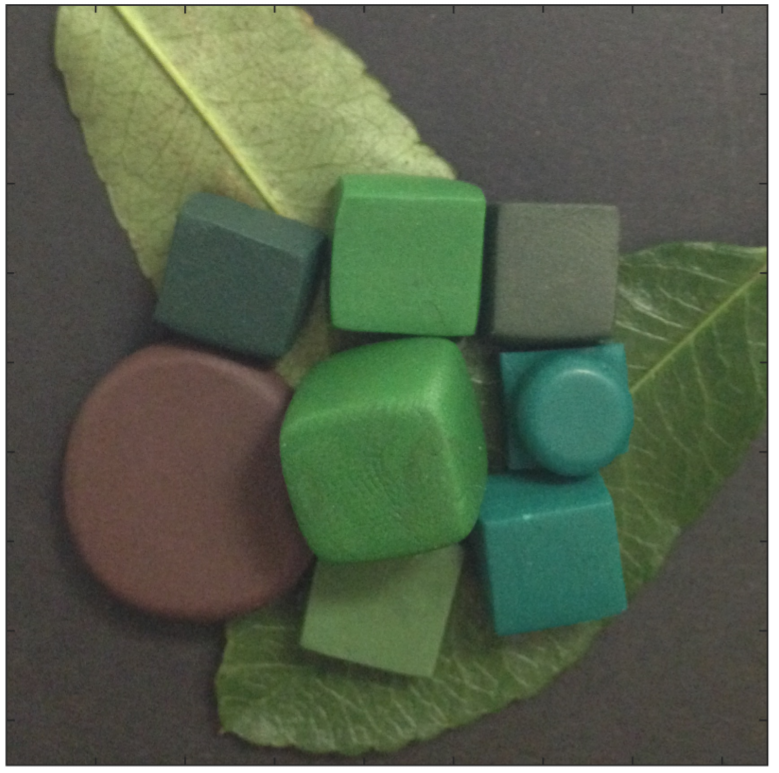}
	}
	\subfloat[]{
		\includegraphics[width=0.3\linewidth]{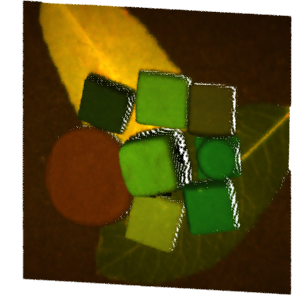}
	}
	\caption{(a) is an RGB image of the target and (b) shows the 3D reconstructed scene (the colors were generated according to the CIE 1931 RGB color space).}
	\label{FIG:blocks_leaves_rgb}
	\vspace{-0.3cm}
\end{figure}

\begin{figure}[!h]
	\centering
	\includegraphics[width=.47\textwidth]{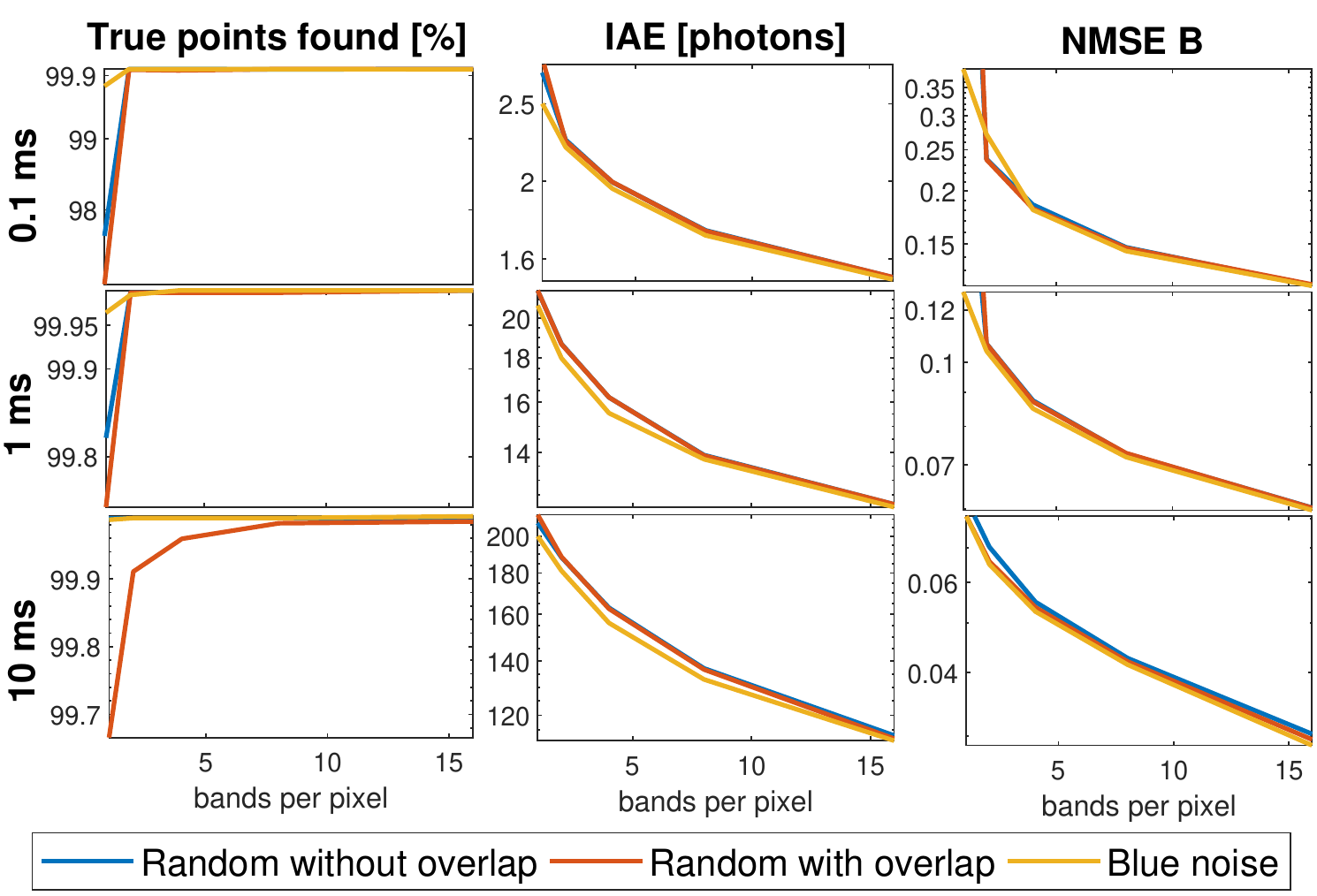}
	\caption{$F_{\textrm{true}}(\tau)$, $F_{\textrm{false}}(\tau)$ and IAE$(\tau)$ obtained with the proposed reconstruction method for different acquisition times and sensed bands per pixel.}
	\label{FIG:results codes}
	\vspace{-0.3cm}
\end{figure}

\begin{figure}[!h]
	\centering
	\includegraphics[width=.7\linewidth]{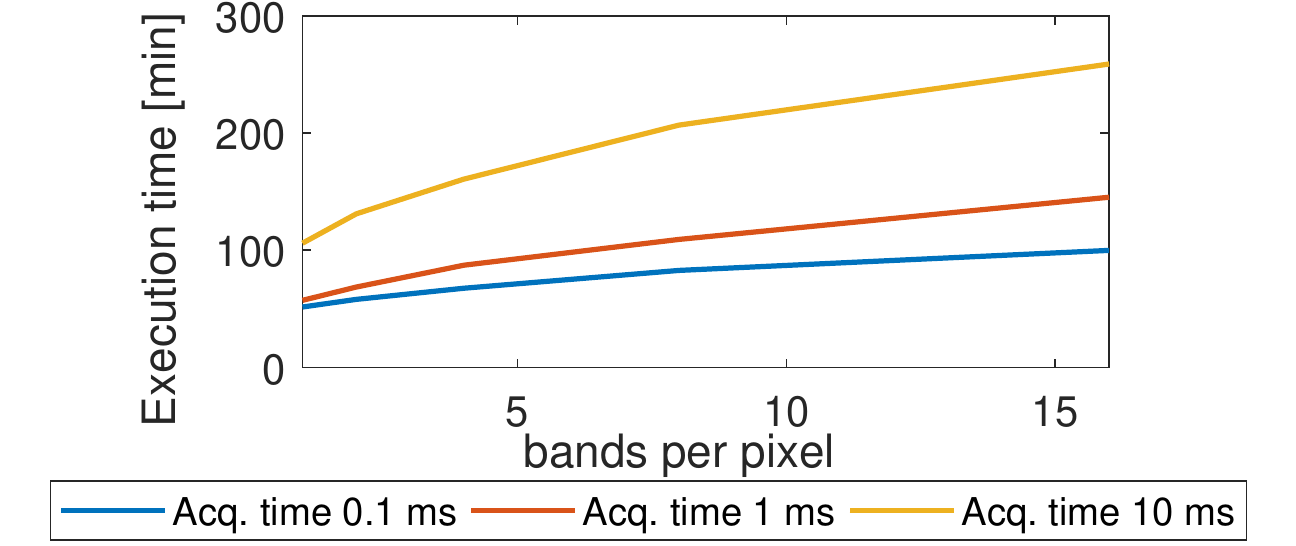}
	\caption{Total execution time for different number of sensed bands per pixel and acquisition times of 0.1, 1 and 10 ms.}
	\label{FIG:exec time}
	\vspace{-0.3cm}
\end{figure}

The proposed subsampling scheme was evaluated on a real MSL dataset \cite{altmann2017robust}. The scene consists of $L=32$ wavelengths sampled at regular intervals of 10 nm from 500 nm to 810 nm, $N_r=N_c=198$ pixels and $T=4500$ histogram bins. The target is composed by a series of blocks of different types of clay and two leaves. \Cref{FIG:blocks_leaves_rgb} shows an RGB image of the scene and the 3D reconstruction \textcolor{black}{using acquisition times up to 10 ms} per wavelength per pixel. We compare the blue noise codes mentioned in \Cref{SEC:CS} with the random schemes introduced in \cite{altmann2017undersample}, all yielding the same total number of measurements and acquisition time:
\begin{enumerate}
	\item Random sampling without overlap: $W$ out of $L$ bands per pixel are sampled without replacement (i.e., for a given pixel, each wavelength is measured at most once).
	\item Random sampling without overlap: For each wavelength, $W/L$\% of the pixels are sampled without replacement.
	\item Proposed sampling method: For each wavelength, $W/L$\% of the pixels are chosen following the scheme presented in \Cref{SEC:CS}.
\end{enumerate}
The codes were evaluated for $W=1,2,4,8,16$ bands per pixel and acquisition times of 0.1, 1 and 10 ms per measurement (i.e., the histogram of one wavelength), using as ground-truth the reconstruction obtained with all the measurements and an acquisition time of 10 ms. \Cref{FIG:results codes} shows the percentage of true detections, IAE and background NMSEs for all codes, acquisition times and numbers of sensed bands per pixel $W$. \textcolor{black}{All the evaluated compressive strategies yield good results, where a small improvement can be obtained by the use of blue noise codes.} In terms of total number of estimated points, the blue noise codes achieve better performance in high compression scenarios $W=1,2$ and low acquisition times (0.1 and 1 ms). For example, for an acquisition time of 1 ms, almost all points are reconstructed using blue noise codes, whereas the random codes only yield around $97\%$ of the ground-truth points. The choice of \textcolor{black}{blue noise codes} has a stronger impact in terms of IAE, achieving smaller IAE for all acquisition times and number of bands per pixel. 
\Cref{FIG:exec time} shows the execution time \textcolor{black}{for acquisition times of 10, 1 and 0.1 ms} and different numbers of sensed bands. The proposed RJ-MCMC sampler has a complexity proportional to the number of photon detections in the support of the impulse response around the 3D point being modified, whereas the background update has a complexity proportional to the total number of active histogram bins in the Lidar scene. The background extraction step required around $15\%$ of the total execution time, which could be significantly reduced if all the bands were processed in parallel instead of sequentially as it is done in the current implementation. \textcolor{black}{All the experiments were performed using a Matlab R2018a implementation on a i7-3.0 GHz desktop computer (16GB RAM)}.

\begin{figure}[!h]
	\centering
	\includegraphics[width=0.7\linewidth]{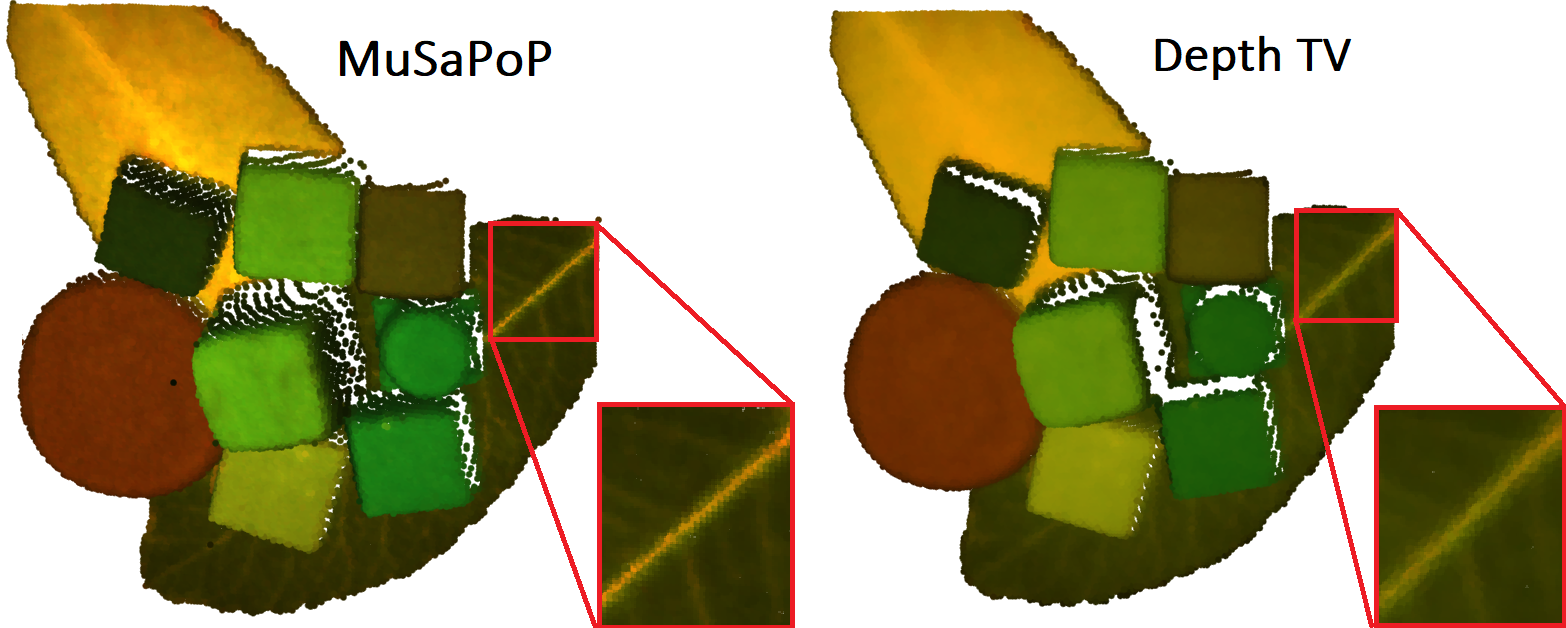}
	\caption{From left to right: 3D reconstructions obtained by the proposed method and Depth TV for an acquisition time of 10 ms. Note that Depth TV tends to smooth out fine scale details (zoom for better visualization). Moreover, the thresholding step used in Depth TV removes some low intensity points in the borders of each 3D object.}
	\label{FIG:no_background_leaves}
	\vspace{-0.3cm}
\end{figure}

\begin{figure}[!h]
	\centering
	\includegraphics[width=.65\linewidth]{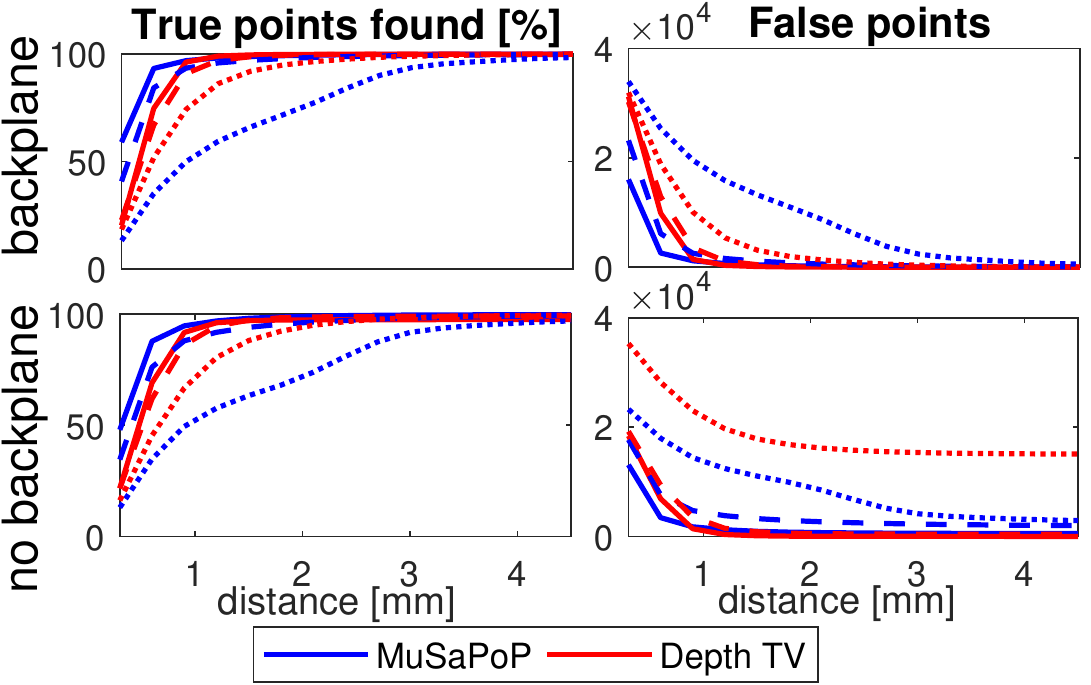}
	\caption{\textcolor{black}{True and false point detections for MuSaPoP and Depth TV for the real MSL dataset with (top row) and without (bottom row) backplane. Solid, dashed and dotted lines represent the datasets with acquisition times of 1, 0.1 and 0.01 ms respectively.}}
	\label{FIG:depthTV}
	\vspace{-0.3cm}
\end{figure}

\textcolor{black}{Finally, we compared the performance of MuSaPoP with the single-depth multiple-wavelength algorithm by Altmann et al. \cite{altmann2017undersample}. The algorithm is referred to as Depth TV and considers total variation regularizations for the background, reflectivity and depth images. Note that this method requires a (global) depth interval where all signal photons are found, which is given manually by the user. We also considered a target detection scenario \textcolor{black}{(i.e., some pixels without surfaces)}, by removing the backplane of the scene and keeping only photons associated with background levels. In this case, we post-process the Depth TV estimates, removing points with a mean normalized intensity below 10\%, which gave the best results across the evaluated datasets. In both experiments, we used the blue noise codes with $W=8$ wavelengths per pixel out of $L=32$. \Cref{FIG:no_background_leaves} shows the 3D reconstructions obtained by MuSaPoP and Depth TV using an acquisition time of 10 ms. \Cref{FIG:depthTV} shows the performance of both algorithms in terms of true and false detections. MuSaPoP performs better in the 1 and 0.1 ms cases, whereas Depth TV obtains better depth estimates in the lowest acquisition time case (0.01 ms). However, in the 0.01 ms case without backplane, the intensity thresholding step does not remove backplane points, hence obtaining a very large number of false detections. This result illustrates the inefficiency of simple thresholding in target detection scenarios, whereas MuSaPoP includes these cases within its general formulation. \Cref{tab:depth_tv} shows the performance of both algorithms in terms of IAE, background NMSE and execution time. The proposed method yields a better IAE than Depth TV (approximately half), as the latter tends to smooth out details within the blocks and leaves, as shown in \Cref{FIG:no_background_leaves}. Moreover, in terms of background NMSE, Depth TV fails to provide good estimates in the low-photon cases, as it only considers photon counts within the global interval without signal returns. The execution time of Depth TV was significantly higher than MuSaPoP.}

\begin{table}[]
	\centering
	\begin{tabular}{|c|c|c|c|c|c|c|c|}
		\hline
		\multicolumn{2}{|c|}{Backplane} & \multicolumn{3}{c|}{Yes} & \multicolumn{3}{c|}{No} \\ \hline
		\multicolumn{2}{|c|}{Acq. time {[}ms{]}} & 1 & 0.1 & 0.01 & 1 & 0.1 & 0.01 \\ \hline
		\multirow{2}{*}{\begin{tabular}[c]{@{}c@{}}IAE\\ {[}photons{]}\end{tabular}} & Depth TV & 36 & 3.7 & 0.5 & 45 & 4.9 & 0.8 \\ \cline{2-8} 
		& MuSaPoP & \textbf{14} & \textbf{1.7} & \textbf{0.3} & \textbf{19} & \textbf{2.4} & \textbf{0.4} \\ \hline
		\multirow{2}{*}{\begin{tabular}[c]{@{}c@{}}Bkg.\\ NMSE\end{tabular}} & Depth TV & 0.12 & \textgreater{}1 & \textgreater{}1 & 0.12 & \textgreater{}1 & \textgreater{}1 \\ \cline{2-8} 
		& MuSaPoP & \textbf{0.04} & \textbf{0.11} & \textbf{0.36} & \textbf{0.04} & \textbf{0.11} & \textbf{0.26} \\ \hline
		\multirow{2}{*}{\begin{tabular}[c]{@{}c@{}}Execution\\ time {[}h{]}\end{tabular}} & Depth TV & 19.8 & 17.9 & 17.7 & 19.3 & 17.9 & 17.7 \\ \cline{2-8} 
		& MuSaPoP & \textbf{2.8} & \textbf{1.4} & \textbf{1.0} & \textbf{1.5} & \textbf{1} & \textbf{0.8} \\ \hline
	\end{tabular}
	\caption{IAE, background NMSE and execution time of Depth TV and the proposed method for the blocks and leaves dataset with and without the backplane.}
	\label{tab:depth_tv}
	\vspace{-0.3cm}
\end{table}

\section{Conclusions and future work}\label{SEC:Conclusions}
This paper has studied a new 3D reconstruction algorithm referred to as MuSaPoP using multispectral Lidar data, which is able to find multiple surfaces per pixel. The proposed method leads to better reconstruction quality than other alternatives, as it considers all measured wavelengths in a single observation model. While based on some ideas initially investigated in ManiPoP \cite{tachella2018manipop}, MuSaPoP also relies on new strategies to deal with the very high dimensionality of the multispectral problem. The first novelty is the use of an empirical Bayes prior for the background levels, which speeds up significantly the RJ-MCMC algorithm. \textcolor{black}{A second improvement is the adapted dilation/erosion and split/merge moves for the multispectral case, profiting from SBR estimates to increase the acceptance rate}. Finally, the subsampling strategy further reduces both the algorithm's complexity and total number of measurements, leading to faster acquisitions and reconstructions. The sparse point cloud representation of the proposed method speeds up the computations proportionally to the number of measurements, whereas other dense models \cite{shin2016computational,halimi2016restoration} would not achieve similar improvements.

Further work will be devoted to the design of compressive systems that are not limited to subsampling the spectral cube. Moreover, the compression can also be extended to depth information, for example using a range-gated camera \cite{Ren18gated}.

\vspace{-0.3cm}
\appendices
\section{Gamma Markov random field}\label{APP:gmrf}
Most classical prior distributions (often referred to as regularization terms in the convex optimization literature) for images, such as Laplacian \cite{rue2005gaussian} and total variation \cite{chambolle_pock_2016}, only penalize variations between neighbouring pixels, ignoring the mean intensity of the image. However, the gamma Markov random field \cite{Dikmen2010}, includes a penalization on the mean intensity, which promotes pixels with smaller values. This can be shown by inspecting the marginal distribution (defined as \cite{tachella2018bay}),
\begin{align}
\label{EQ:marginalized GMRF}
p(\mat{B} |\alpha_{B}) &\propto \prod_{\ell=1}^{L} \prod_{i=1}^{N_c} \prod_{j=1}^{N_r} \frac {b_{i,j,\ell}^{\alpha_{B}-1}} { \left(\sum_{(i',j')\in \mathcal{M}_B} b_{i',j',\ell}\right)^{\alpha_{B}}}
\end{align}
where $\alpha_{B}$ is a hyperparameter controlling the degree of smoothness and $\mathcal{M}_B$ denotes the set of pixels in the neighbourhood of pixel $(i,j)$. For an image of constant intensity $c$, we have  $b_{i',j',\ell} = b_{i,j,\ell} = c$ for all pixels and spectral bands, yielding the density
\begin{equation}
p(\mat{B} |\alpha_{B}) \propto \prod_{\ell=1}^{L}\prod_{i=1}^{N_c} \prod_{j=1}^{N_r} c^{-1} =
 \prod_{i=1}^{N_c} \prod_{j=1}^{N_r} c^{-L}.
\end{equation}
This dependency on the mean promotes reconstructions with lower background levels, decreasing the acceptance ratio of death and erosion moves (that propose to increase the background levels). In the case of ManiPoP, only one band is considered ($L=1$). Thus, the bias towards smaller background levels does not impact the overall reconstruction significantly. However, in the MSL case ($L\gg1$), the reconstruction quality is reduced, hindering the use of gamma Markov random fields.
\section{Empirical prior for the background levels}\label{APP:KL_div}
The prior for the background levels is chosen to minimize the Kullback-Leibler divergence in \eqref{EQ:KL_div}, where the correlated model $q$ is given by \eqref{EQ:poisson-gaussian}. The minimization of \eqref{EQ:KL_div} can be written as 
\begin{align}
(\mat{K},\mat{\Theta}) = \argmin_{(\mat{K},\mat{\Theta})}  -\mathbb{E}_q\{\log p(\mat{B}|\mat{K},\mat{\Theta})\}.
\end{align}
Considering the product of independent gamma distributions in \eqref{EQ:independent_gamma}, the problem reduces to
\begin{multline}
\label{EQ:simple KL}
(k_{i,j,\ell},\theta_{i,j,\ell}) = \argmin_{(k_{i,j,\ell},\theta_{i,j,\ell})} \frac{\mathbb{E}_q\{b_{i,j,\ell}\}}{\theta_{i,j,\ell}} \\- k_{i,j,\ell}\left(\mathbb{E}_q\{\log b_{i,j,\ell}\}-\log \theta_{i,j,\ell}\right)+\log \Gamma(k_{i,j,\ell})
\end{multline}
for all pixels $i=1,\dots,N_r$ and $j=1,\dots,N_c$ and wavelengths $\ell=1,\dots,L$. The expected values $\mathbb{E}_q\{b_{i,j,\ell}\}$ and $\mathbb{E}_q\{\log b_{i,j,\ell}\}$ cannot be obtained in closed form for the Poisson-Gaussian model of \eqref{EQ:poisson-gaussian}. Thus, we approximate them numerically by obtaining MCMC samples of $\tilde{b}_{i,j,\ell}$. As explained in \cite{tachella2018bay},  off-the-shelf sampling strategies (e.g., Hamiltonian Monte Carlo \cite[Chapter~9]{brooks2011handbook}) do not scale well with the dimension of the problem, being inefficient when applied to large multispectral Lidar datasets. Hence, we consider proposals from a Gaussian approximation of \eqref{EQ:poisson-gaussian} (as detailed in \cite{rue2005gaussian}) using the perturbation optimization algorithm \cite{gilavertperturb2014}, accepting or rejecting them according to the Metropolis-Hastings rule \cite{brooks2011handbook,rue2005gaussian}.
We generate $10^3$ samples $\{\tilde{b}^{(s)}_{i,j,\ell}, s=1,\dots,10^3\}$ and compute the desired expected values as 
\begin{align}
\mathbb{E}_q\{b_{i,j,\ell}\} &= \sum_{s} \exp{\tilde{b}^{(s)}_{i,j,\ell}} \\
\mathbb{E}_q\{\log b_{i,j,\ell}\} &= \sum_{s} \tilde{b}^{(s)}_{i,j,\ell}.
\end{align}
Finally, the values of the hyperparameters are obtained by setting $\theta_{i,j,\ell}=\mathbb{E}_q\{b_{i,j,\ell}\}$ and minimizing \eqref{EQ:simple KL} with a one-dimensional Newton method.
\section{Complete expressions of the RJ-MCMC moves}\label{APP:moves}
The birth move of point $(\myvec{c}_{N_\Phi+1},\mat{r}_{N_\Phi+1})$ has an acceptance ratio given by $\rho=\min\{1,r\left(\myvec{\theta},\myvec{\theta'}\right)\}$ with
\begin{equation*}
r\left(\myvec{\theta},\myvec{\theta'}\right) = \left\{
\begin{array}{ll}
C_1  & \mbox{if } |t_{N_\Phi+1}-t_{n}| > d_{\min} \quad \forall n\ne{N_\Phi+1}:\\
&  x_n=x_{N_\Phi+1} \text{ and } y_n=y_{N_\Phi+1} \\
0 & \mbox{otherwise}
\end{array}
\right.
\end{equation*}
where $C_1$ is defined as
\begin{multline*}
C_1 = \prod_{\ell=1}^{L}\prod_{t=1}^{T}\left(\frac{  \sum_{\substack{n:x_n=i\\ y_n=j \\}}r'_{n,\ell}h_{\ell}(t-t'_{n})+ b'_{i,j,\ell}  }{ \sum_{\substack{n:x_n=i\\ y_n=j \\}}r_{n,\ell}h_{\ell}(t-t_{n})+ b_{i,j,\ell}   } \right)^{z_{i,j,t,\ell}} \frac{p_{\textrm{death}}}{p_{\textrm{birth}}}
\\  \lambda_{a}\gamma_{a}^{-m\left(S(\myvec{c}_{N_\Phi+1}) \setminus  \bigcup_{n'\in \mathcal{M}_{pp}(\myvec{c}_{N_\Phi+1})}S(\myvec{c}_{n'}) \right)}
\frac{1}{N_{\Phi}+1}
\left(\frac{|\mat{P'}|}{|\mat{P}|}  \frac{1}{2\pi\sigma^2}\right)^{\frac{L}{2}}
\\ \prod_{\ell=1}^{L}\exp\left(-\sum_{n'\in\mathcal{M}_{pp}(\myvec{c}_n)}\frac{(m_{N_\Phi+1,\ell}-m_{n',\ell})^2}{2\sigma^2d(\myvec{c}_{N_\Phi+1};\myvec{c}_{n'})}-\frac{{m_{N_\Phi+1,\ell}}^2\beta}{2\sigma^2}\right)
\\
(1-u)^{-L} \prod_{\ell=1}^{L} \exp\left(g_{i,j,\ell}r_{N_\Phi+1,\ell} (1-{w_{\ell}}^{-1})\left(\sum_{t=1}^{T}h_{\ell}(t) \right)\right)
\\ 
\prod_{(i,j)\in \mathcal{M}_B(b_{i,j})} \prod_{\ell=1}^{L} \left(\frac{{b'}_{i,j,\ell}}{{b}_{i,j,\ell}}\right)^{r_{i,j,\ell}-1} \exp(\frac{{b}_{i,j,\ell}-{b'}_{i,j,\ell}}{\theta_{i,j,\ell}}) .
\end{multline*}
Similarly, the death move is accepted with probability $\rho=\min\{1,C_1^{-1}\}$, where the term $\frac{1}{N_{\Phi}+1}$ in the second line is replaced by $\frac{1}{N_{\Phi}}$. The dilation move of point $(\myvec{c}_{N_\Phi+1},r_{N_\Phi+1})$ is accepted with probability 
$\rho=\min\{1,r\left(\myvec{\theta},\myvec{\theta'}\right)\}$ with
\begin{equation*}
r\left(\myvec{\theta},\myvec{\theta'}\right) = \left\{
\begin{array}{ll}
C_2  & \mbox{if } |t_{N_\Phi+1}-t_{n}| > d_{\min} \quad \forall n\ne{N_\Phi+1} \\
&  x_n=x_{N_\Phi+1} \text{ and } y_n=y_{N_\Phi+1} \\
0 & \mbox{otherwise}
\end{array}
\right.
\end{equation*}
where $C_2$ is defined as
\begin{multline*}
C_2 = \prod_{\ell=1}^{L}\prod_{t=1}^{T}\left(\frac{  \sum_{\substack{n:x_n=i\\ y_n=j \\}}r'_{n,\ell}h_{\ell}(t-t'_{n})+ b'_{i,j,\ell}  }{ \sum_{\substack{n:x_n=i\\ y_n=j \\}}r_{n,\ell}h_{\ell}(t-t_{n})+ b_{i,j,\ell}   } \right)^{z_{i,j,t,\ell}}
\\  \frac{p_{\textrm{erosion}}}{p_{\textrm{dilation}}}
\\
\lambda_{a}\gamma_{a}^{-m\left(S(\myvec{c}_{N_\Phi+1}) \setminus  \bigcup_{n'\in \mathcal{M}_{pp}(\myvec{c}_{N_\Phi+1})}S(\myvec{c}_{n'}) \right)}
\left(\frac{|\mat{P'}|}{|\mat{P}|}  \frac{1}{2\pi\sigma^2}\right)^{\frac{L}{2}}
\\
\frac{N_\Phi(2N_b+1)}{\sum_{m\in\mathcal{M}_{pp}(\myvec{c}_{N_\Phi+1})}  \card\mathcal{M}_{pp}(\myvec{c}_{m})} \times
\frac{1}{\sum_{m=1}^{N_\Phi+1} \indicator{\mathbb{Z}_{+}}{\card\mathcal{M}_{pp}(\myvec{c}_{m})}}
\\ \prod_{\ell=1}^{L}\exp\left(-\sum_{n'\in\mathcal{M}_{pp}(\myvec{c}_n)}\frac{(m_{N_\Phi+1,\ell}-m_{n',\ell})^2}{2\sigma^2d(\myvec{c}_{N_\Phi+1};\myvec{c}_{n'})}-\frac{{m_{N_\Phi+1,\ell}}^2\beta}{2\sigma^2}\right)
\\
(1-u)^{-L} \prod_{\ell=1}^{L} \exp\left(g_{i,j,\ell}r_{N_\Phi+1,\ell} (1-{w_{\ell}}^{-1})\left(\sum_{t=1}^{T}h_{\ell}(t) \right) \right)
\\
\prod_{(i,j)\in \mathcal{M}_B(b_{i,j})} \prod_{\ell=1}^{L} \left(\frac{{b'}_{i,j,\ell}}{{b}_{i,j,\ell}}\right)^{r_{i,j,\ell}-1} \exp(\frac{{b}_{i,j,\ell}-{b'}_{i,j,\ell}}{\theta_{i,j,\ell}}).
\end{multline*}
A shift of the point $(\myvec{c}_n,\myvec{r}_n)$ to the new position $\myvec{c}_n'=[x_n,y_n,t'_n]^T$ has an acceptance probability of $\rho=\min\{1,r\left(\myvec{\theta},\myvec{\theta'}\right)\}$ with
\begin{equation*}
r\left(\myvec{\theta},\myvec{\theta'}\right) = \left\{
\begin{array}{ll}
C_3  & \mbox{if }  |t'_{n}-t_{m}| > d_{\min} \quad \forall n\ne m  \\
& x_m=x_n \text{ and } y_m=y_n   \\
0 & \mbox{otherwise}
\end{array}
\right.
\end{equation*}
where  
\begin{multline*}
C_3 = \prod_{\ell=1}^{L}\prod_{t=1}^{T}\left(\frac{  \sum_{\substack{n:x_n=i\\ y_n=j \\}}r'_{n,\ell}h_{\ell}(t-t'_{n})+ b'_{i,j,\ell}  }{ \sum_{\substack{n:x_n=i\\ y_n=j \\}}r_{n,\ell}h_{\ell}(t-t_{n})+ b_{i,j,\ell}   } \right)^{z_{i,j,t,\ell}}
\\
\prod_{\ell=1}^{L}\exp\left(-\frac{1}{2\sigma^2}\left(\sum_{n'\in\mathcal{M}_{pp}(\myvec{c'}_n)}\frac{(m_{n,\ell}-m_{n',\ell})^2}{d(\myvec{c'}_{n};\myvec{c}_{n'})}\right)\right)
\\
\left(\frac{|\mat{P'}|}{|\mat{P}|}\right)^{\frac{L}{2}}\prod_{\ell=1}^{L}\exp\left(\frac{1}{2\sigma^2}\left(\sum_{n'\in\mathcal{M}_{pp}(\myvec{c}_n)}\frac{(m_{n,\ell}-m_{n',\ell})^2}{d(\myvec{c}_{n};\myvec{c}_{n'})}\right)\right) 
\\
\gamma_{a}^{-m\left(S(\myvec{c'}_{n}) \setminus  \bigcup_{n'\in \mathcal{M}_{pp}(\myvec{c'}_{n})}S(\myvec{c}_{n'}) \right)+m\left(S(\myvec{c}_{n}) \setminus  \bigcup_{n'\in \mathcal{M}_{pp}(\myvec{c}_{n})}S(\myvec{c}_{n'}) \right)}.
\end{multline*}
A mark update randomly picks a point $(\myvec{c}_n,\myvec{r}_n)$ and proposes a new spectral signature $\myvec{r}'_{n}=\log(\myvec{m}'_{n})$. Each spectral log-intensity is accepted independently with probability $\rho=\min\{1,C_4\}$, where 
\begin{multline*}
C_4 =  \prod_{\ell=1}^{L}\prod_{t=1}^{T}\left(\frac{  \sum_{\substack{n:x_n=i\\ y_n=j \\}}r'_{n,\ell}h_{\ell}(t-t'_{n})+ b'_{i,j,\ell}  }{ \sum_{\substack{n:x_n=i\\ y_n=j \\}}r_{n,\ell}h_{\ell}(t-t_{n})+ b_{i,j,\ell}   } \right)^{z_{i,j,t,\ell}}
\\ \exp\left(-\frac{1}{2\sigma^2}\left(\sum_{n'\in\mathcal{M}_{pp}(\myvec{c'}_n)}\frac{(m'_{n,\ell}-m_{n',\ell})^2}{d(\myvec{c'}_{n};\myvec{c}_{n'})}+{m'_{n,\ell}}^2\beta\right)\right) 
\\ \exp\left(\frac{1}{2\sigma^2}\left(\sum_{n'\in\mathcal{M}_{pp}(\myvec{c}_n)}\frac{(m_{n,\ell}-m_{n',\ell})^2}{d(\myvec{c}_{n};\myvec{c}_{n'})}+m_{n,\ell}^2\beta\right)\right)
\\
\prod_{\ell=1}^{L} \exp\left(g_{i,j,\ell} (r_{n,\ell}-r'_{n,\ell}) (1-{w_{\ell}}^{-1})\left(\sum_{t=1}^{T}h_{\ell}(t) \right)\right).
\end{multline*}
The split move from $(\myvec{c}_n=[x_n,y_n,t_{n}]^T,\myvec{r}_n)$ to $(\myvec{c'}_{k_1}=[x_n,y_n,t'_{k_1}]^T,\myvec{r}'_{k_1})$ and $(\myvec{c'}_{k_2}=[x_n,y_n,t'_{k_2}]^T,\myvec{r}'_{k_2})$ is accepted with probability $\rho=\min\{1,r\left(\myvec{\theta},\myvec{\theta'}\right)\}$, where
\begin{equation*}
r\left(\myvec{\theta},\myvec{\theta'}\right) = \left\{
\begin{array}{ll}
C_5  & \mbox{if }  |t'_{n}-t_{m}| > d_{\min} \quad \forall n\ne m: \\
& x_m=x_n \text{ and } y_m=y_n   \\
0 & \mbox{otherwise}
\end{array}
\right.
\end{equation*}
and
\begin{multline*}
C_5 = \prod_{\ell=1}^{L}\prod_{t=1}^{T}\left(\frac{  \sum_{\substack{n:x_n=i\\ y_n=j \\}}r'_{n,\ell}h_{\ell}(t-t'_{n})+ b'_{i,j,\ell}  }{ \sum_{\substack{n:x_n=i\\ y_n=j \\}}r_{n,\ell}h_{\ell}(t-t_{n})+ b_{i,j,\ell}   } \right)^{z_{i,j,t,\ell}} 
\\
(u(1-u))^{-L} N_\Phi (\card\text{points in } \mat{\Phi} \text{ that verify  \eqref{EQ: merge_condition}})^{-1} 
\left(\frac{|\mat{P'}|}{|\mat{P}|}\right)^{\frac{L}{2}}
\\
\prod_{\ell=1}^{L}\exp\left(-\frac{1}{2\sigma^2}\left(\sum_{n'\in\mathcal{M}_{pp}(\myvec{c'}_{k_1})}\frac{(m_{k_1,\ell}-m_{n',\ell})^2}{d(\myvec{c'}_{k_1};\myvec{c}_{n'})}\right)\right) 
\\
\prod_{\ell=1}^{L}\exp\left(-\frac{1}{2\sigma^2}\left(\sum_{n'\in\mathcal{M}_{pp}(\myvec{c'}_{k2})}\frac{(m_{k_1,\ell}-m_{n',\ell})^2}{d(\myvec{c'}_{k_2};\myvec{c}_{n'})}\right)\right) 
\\
\prod_{\ell=1}^{L}\exp\left(\frac{1}{2\sigma^2}\left(\sum_{n'\in\mathcal{M}_{pp}(\myvec{c}_n)}\frac{(m_{n,\ell}-m_{n',\ell})^2}{d(\myvec{c}_{n};\myvec{c}_{n'})}\right)\right) 
\\
\gamma_{a}^{-m\left(S(\myvec{c'}_{k_1}) \setminus  \bigcup_{n'\in \mathcal{M}_{pp}(\myvec{c'}_{k_1})}S(\myvec{c}_{n'}) \right)+m\left(S(\myvec{c}_{n}) \setminus  \bigcup_{n'\in \mathcal{M}_{pp}(\myvec{c}_{n})}S(\myvec{c}_{n'}) \right)}
\\
\lambda_{a}\gamma_{a}^{-m\left(S(\myvec{c'}_{k_2}) \setminus  \bigcup_{n'\in \mathcal{M}_{pp}(\myvec{c'}_{k_2})}S(\myvec{c}_{n'}) \right)}2(d_{\min}+L_h) \frac{p_{\textrm{merge}}}{p_{\textrm{split}}}.
\end{multline*}
Finally, the merge move is accepted with probability $\rho=\min\{1,C_5^{-1}\}$.

\section*{Acknowledgment}
We would like to thank the single-photon group\footnote{http://www.single-photon.com} led by Prof. G. S. Buller for providing us the real MSL dataset used in this paper.
\ifCLASSOPTIONcaptionsoff
  \newpage
\fi



%
\bibliography{MuSaPoP}
\bibliographystyle{IEEEtran}

\end{document}